\RequirePackage{ifpdf}
\documentclass[11pt,letterpaper]{article}
\pdfoutput=1
\usepackage{jheppub}
\usepackage{amsmath,amssymb,amsfonts,bm,amscd}
\usepackage{graphicx}
\usepackage{subcaption}
\usepackage{verbatim}
\usepackage{fancybox}
\usepackage{changepage}
\usepackage{slashed}
\usepackage{url}
\usepackage{array}
\usepackage{hhline}
\usepackage{tabularx}
\usepackage{tikz}

\usepackage{rotating,booktabs}
\usepackage[verbose]{placeins}
\usepackage{blindtext}

\graphicspath {{figures/}}

\newcommand{\bea}{\begin{eqnarray}}
\newcommand{\eea}{\end{eqnarray}}
\newcommand{\be}{\begin{equation}}
\newcommand{\ee}{\end{equation}}

\newcommand{\Z}{{\mathbb Z}}
\newcommand{\R}{{\mathbb R}}
\newcommand{\C}{{\mathbb C}}

\newcommand{\fr}{{\frak r}}
\newcommand{\pbra}[1]{\left(#1\right)}

\newcommand{\dbra}[1]{[\![ #1 ]\!] }
\newcolumntype{x}[1]{>{\centering\arraybackslash}p{#1}}

\def\eg{{\textit{e.g.}}}
\def\ie{{\textit{i.e.}}}

\def\ft{\frak{t}} 

\def\frak{\mathfrak}

\def\tilde{\widetilde}
\def\hat{\widehat}
\def\bar{\overline}

\def\CB{{\mathcal B}}
\def\CC{{\mathcal C}}

\def\CH{{\mathcal H}}
\def\CI{{\mathcal I}}

\def\CL{{\mathcal L}}
\def\CM{{\mathcal M}}
\def\CN{{\mathcal N}}
\def\CO{{\mathcal O}}
\def\CP{{\mathcal P}}

\def\CS{{\mathcal S}}
\def\CT{{\mathcal T}}

\def\CZ{{\mathcal Z}}

\newcommand{\cp}{{\mathbb{C}}{\mathbf{P}}}
\renewcommand{\bar}{\overline}
\renewcommand{\hat}{\widehat}

\def\^{{\wedge}}
\def\*{{\star}}

\newcommand{\beq}{\begin{equation}\begin{aligned}}
\newcommand{\eeq}{\end{aligned}\end{equation}}

\title{Equivariant Verlinde algebra from superconformal index and Argyres-Seiberg duality}

\author[a,b]{Sergei Gukov,}
\author[a,c]{Du Pei,}
\author[d,a,e,f]{Wenbin Yan,}
\author[a]{and Ke Ye}

\affiliation[a]{Walter Burke Institute for Theoretical Physics,
California Institute of Technology, \\Pasadena, CA 91125, USA}

\affiliation[b]{Max-Planck-Institut f\"ur Mathematik, Vivatsgasse 7, D-53111 Bonn, Germany}

\affiliation[c]{Center for Quantum Geometry of Moduli Spaces, Department of Mathematics, University of Aarhus, \\DK-8000, Denmark}

\affiliation[d]{Yau Mathematical Sciences Center, Tsinghua University, Beijing, 100084, China}

\affiliation[e]{Center for Mathematical Sciences and Applications, Harvard University, \\Cambridge, MA 02138, USA}

\affiliation[f]{Jefferson Physical Laboratory, Harvard University, Cambridge, MA 02138, USA}

\emailAdd{gukov@theory.caltech.edu}
\emailAdd{pei@caltech.edu}
\emailAdd{wenbin.yan@gmail.com}
\emailAdd{kye@caltech.edu}

\abstract{In this paper, we show the equivalence between two seemingly distinct 2d TQFTs: one comes from the ``Coulomb branch index'' of the class ${\cal S}$ theory $T[\Sigma,G]$ on $L(k,1) \times S^1$, the other is the $^L\! G$ ``equivariant Verlinde formula'', or equivalently partition function of $^L\! G_{\mathbb{C}}$ complex Chern-Simons theory on $\Sigma\times S^1$. We first derive this equivalence using the M-theory geometry and show that the gauge groups appearing on the two sides are naturally $G$ and its Langlands dual $^L\! G$. When $G$ is not simply-connected, we provide a recipe of computing the index of $T[\Sigma,G]$ as summation over the indices of $T[\Sigma,\tilde{G}]$ with non-trivial background 't Hooft fluxes, where $\tilde{G}$ is the universal cover of $G$. Then we check explicitly this relation between the Coulomb index and the equivariant Verlinde formula for $G=SU(2)$ or $SO(3)$. In the end, as an application of this newly found relation, we consider the more general case where $G$ is $SU(N)$ or $PSU(N)$ and show that equivariant Verlinde algebra can be derived using field theory via (generalized) Argyres-Seiberg duality. We also attach a Mathematica notebook that can be used to compute the $SU(3)$ equivariant Verlinde coefficients.
\\
\\
\\
\\
\\
\\
\\
{\tt CALT-TH-2016-012}
}

\begin{document}

\maketitle

\section{Introduction}

The complex Chern-Simons theory was studied by embedding it into string theory in \cite{equivariant}, and the starting point is the following configuration of M-theory fivebranes that is often used to study the 3d-3d correspondence~\cite{DGH, Terashima:2011qi, Dimofte:2011ju, Dimofte:2011py,Cecotti:2011iy}:
\beq
\begin{matrix}
{\mbox{\rm space-time:}} & \qquad & L(k,1)_b & \times & T^* M_3 & \times & \R^2\\
& \qquad &  &  & \cup \\
N~{\mbox{\rm fivebranes:}} & \qquad & L(k,1)_b & \times & M_3 
\end{matrix}
\label{3d3d1}
\eeq
If one reduces along the squashed Lens space $L(k,1)_b$, one obtains complex Chern-Simons theory at level $k$ on $M_3$ \cite{Cordova:2013cea}. Even in the simple case where $M_3$ is the product of a Riemann surface $\Sigma$ with a circle $S^1$, this system is extremely interesting and can be used to gain a lot of insight into complex Chern-Simons theory. For example, the partition function of the 6d $(2,0)$-theory on this geometry gives the ``equivariant Verlinde formula'', which can be identified with the dimension of the Hilbert space of the complex Chern-Simons theory at level $k$ on $\Sigma$:
\be\label{EVF}
Z_{\text{M5}}(L(k,1)\times \Sigma \times S^1, \beta)=\dim_\beta \CH_{\text{CS}}(\Sigma,k).
\ee
Here $\beta$ is an ``equivariant parameter'' associated with a geometric $U(1)_\beta$ action whose precise definition will be reviewed in section \ref{sec:EVACBI}. The left-hand side of \eqref{EVF} has been computed in several ways in \cite{equivariant} and \cite{appetizer}, each gives unique insight into the equivariant Verlinde formula, the complex Chern-Simons theory and the 3d-3d correspondence in general. In this paper, we will add to the list yet another method of computing the partition of the system of M5-branes by relating it to superconformal indices of class $\CS$ theories.

The starting point is the following observation. For $M_3=\Sigma\times S^1$, the setup \eqref{3d3d1} looks like:
\beq
\begin{matrix}
{\mbox{$N$ fivebranes:}} & \qquad & L(k,1)_b & \times & \Sigma & \times &S^1 \\
& \qquad &   &  & \cap \\
{\mbox{space-time:}} & \qquad & L(k,1)_b & \times & T^* \Sigma & \times & S^1 & \times & \R^3
\end{matrix}
\label{3d3d2}
\eeq
and it is already very reminiscent of the setting of Lens space superconformal indices of class $\CS$ theories \cite{Romelsberger:2005eg, Kinney:2005ej, Gadde:2011uv, Alday:2013rs, Razamat:2013jxa}:
\beq
\begin{matrix}
{\mbox {\textrm{$N$ fivebranes:}}} & \qquad & L(k,1)\times S^1& \times & \Sigma &  & \\
& \qquad &   &  & \cap \\
{\mbox{\rm space-time:}} & \qquad & L(k,1)\times S^1 & \times & T^* \Sigma &  &  & \times & \R^3\\
& \qquad &  \!\!\!\!\!\!\!\!\!\!\!\!\circlearrowright &  &\!\!\!\!\!\!\! \circlearrowright &  &  &  &  \circlearrowright \\
{\mbox{\rm symmetries:}} & \qquad & \!\!\!\!\!\! SO(4)_E &   & \!\!\!\! U(1)_N &   &   &   & SU(2)_R
\end{matrix}.
\label{IndGeo}
\eeq
In this geometry, one can turn on holonomies of the symmetries along the $S^1$ circle in a supersymmetric way and introduce three ``universal fugacities'' $(p,q,t)$. Then the partition function of M5-branes in this geometry is the Lens space superconformal index of the 4d $\CN=2$ theory $T[\Sigma]$ of class $\CS$:
\be\label{Index}
Z_{\text{M5}}(L(k,1)\times S^1\times \Sigma, p,q,t)={\CI}(T[\Sigma],p,q,t),
\ee
where we have adopted the following convention for the index\footnote{In the literature there are several other conventions in use. The other two most commonly used conventions for universal fugacities are $(\rho, \sigma, \tau)$ which are related to our convention via $p = \sigma \tau, q = \rho \tau, t = \tau^2$, and $(t,y,v)$ with $t = \sigma^{\frac{1}{6}}\rho^{\frac{1}{6}}\tau^{\frac{1}{3}}, y = \sigma^{\frac{1}{2}} \rho^{-\frac{1}{2}}, v = \sigma^{\frac{2}{3}}\rho^{\frac{2}{3}}\tau^{-\frac{2}{3}}$.}
\be
{\CI} (p,q,t) = {\rm{Tr}} (-1)^F p^{\frac{1}{2} \delta_{1+}}q^{\frac{1}{2} \delta_{1-}}t^{R+r} e^{-\beta''  {\tilde \delta}_{1\dot{-}}}.
\label{4d index}
\ee
As the left-hand sides of \eqref{EVF} and \eqref{Index} are closely related, it is very tempting to ask whether the equivariant Verlinde formula for a Riemann surface $\Sigma$, parametrized by $\beta\in\R$, can actually be embedded as a one-parameter family inside the three-parameter space of superconformal indices of the theory $T[\Sigma]$. 
The goal of this paper is to give strong evidence for the following proposal
\be\label{Statement}
\boxed{\genfrac{}{}{0pt}{}{\text{Equivariant Verlinde formula}}{\text{ at level $k$ on $\Sigma$ for group $G$}}} \quad = \quad \boxed{\genfrac{}{}{0pt}{}{\text{Coulomb branch index}}{\text{of $T[\Sigma,^L\!G]$ on $L(k,1)\times S^1$}}}\, ,
\ee
where the Coulomb branch index is the one-parameter family obtained by taking $p,q,t\rightarrow 0$ while keeping $\ft=pq/t$ fixed. 

To clarify the proposed relation \eqref{Statement}, we first give a few remarks:
\begin{enumerate}
	\item When we fixed $\Sigma$, $G$ and $k\in\Z$, both sides depend on a real parameter and the identification between them is given by $\ft=e^{-\beta}$.
	\item We will assume $\frak{g}=\mathrm{Lie}\,G$ is of type ADE (modulo possible abelian factors), as $T[\Sigma,^L\!G]$, with $^L\!G$ being the Langlands dual group of $G$, is not yet defined in the literature when $\frak{g}$ is not simply-laced. Then we have $\frak{g}=^L\!\frak{g}$.
	\item When $G$ is simple but not simply-connected, the left-hand side of \eqref{Statement} is only defined when $k$ annihilates $\pi_1(G)$ (under the natural $\Z$-action on this abelian group), and the proposal is meant for these values of $k$.
	\item When $^L\!G$ is simple but not simply-connected, the theory $T[\Sigma,^L\!G]$ is not yet defined. Denote the universal cover of $^L\!G$ (which equals the universal cover of $G$ as $\frak{g}$ is of type ADE) as $\tilde{G}$. We will interpret the Coulomb index of $T[\Sigma,^L\!G]$ as a summation of indices of $T[\Sigma,\tilde{G}]$ with insertion of all possible 't Hooft fluxes valued in $\pi_1(^L\!G)$. The insertion is along the 2d surface $S^1\times S^1_{\text{Hopf}}\subset S^1\times L(k,1)$, where $S^1_{\text{Hopf}}$ is the Hopf fiber of the Lens space $L(k,1)$.\footnote{Another natural definition of the partition function of $T[\Sigma,^L\!G]$ is as the summation over only fluxes valued in $H^2(L(k,1),\pi_1(^L\!G))=\Z_k\otimes\pi_1(^L\!G)$, which is a subgroup of $\pi_1(^L\!G)$. If one takes this as the definition, then \eqref{Statement} is correct when $k$ also annihilates $\pi_1(^L\!G)$.} We will give a concrete argument in Section~\ref{fluxargument} using string theory for the $A_{N-1}$ series by starting with $\frak{g}=\frak{u}(N)$, and show that this summation naturally arises when we decouple the abelian $\frak{u}(1)$ factor.
	\item Conceptually, the reason why $G$ appears on the left of \eqref{Statement} while $^L\!G$ appears on the right can be understood as follows. The left-hand side of \eqref{Statement} can be viewed as certain B-model partition function of the Hitchin moduli space $\CM_H(\Sigma,G)$ \cite{Hitchin:1986vp}. Mirror symmetry will produce the Hitchin moduli space associated with the dual group $\CM_H(\Sigma,^L\!G)$ \cite{hausel2003mirror, Donagi:2006cr}, and as we will argue in later sections, the corresponding A-model partition function of $\CM_H(\Sigma,^L\!G)$ can be identified with the right-hand side of \eqref{Statement}. 
	
\end{enumerate}

To further illustrate \eqref{Statement}, we will present the simplest example where $k=1$ and $G$ is simply connected. The equivariant Verlinde formula can be obtained using the TQFT structure studied in \cite{Andersen:2016hoj}
\be\label{EVFk=1}
\dim_\beta \CH_{\text{CS}}(\Sigma,G_{\mathbb{C}},k=1)=\frac{|\CZ(G)|^g}{\left[\prod_{i=1}^{\mathrm{rank}\,G}(1-\ft^{d_i})^{h_i}\right]^{g-1}} \ ,
\ee
where $|\CZ(G)|$ is the order of the center of group $G$, $d_i$'s are degrees of the fundamental invariants of $\frak{g}=\mathrm{Lie}\,G$, and $h_i$'s are the dimension of the space of $d_i$-differentials on $\Sigma$. The reader may have already recognized that \eqref{EVFk=1} is exactly the Coulomb branch index of $T[\Sigma,G]$ on $L(k=1,1)=S^3$ times $|\CZ(G)|^g$. As we will explain in great detail later, the $|\CZ(G)|^g$ factor comes from summation over 't Hooft fluxes, which are labeled precisely by elements in $\CZ(G)\simeq \pi_1(^L\!G)$. The $g$ power morally originates from the fact that there are $g$ ``independent gauge nodes'' in the theory $T[\Sigma,G]$ (\textit{i.e.}~one copy of $G$ for each handle of $\Sigma$). So \eqref{EVFk=1} agrees with the Coulomb index of $T[\Sigma,^L\!G]$. 

For $k>1$, the relation \eqref{Statement} becomes more non-trivial, and each flux sector gives generally different contribution. Even if one sets $\ft=0$, the identification of Verlinde algebra with the algebra of allowed 't Hooft fluxes in $T[\Sigma,G]$ is novel.

This paper is organized as follows. In section \ref{sec:EVACBI}, we examine more closely the two fivebranes systems \eqref{3d3d1} and \eqref{IndGeo}, and give arguments supporting the relation \eqref{Statement} between the equivariant Verlinde formula and the Coulomb branch index. In section \ref{sec:SU2}, after reviewing basic facts and ingredients of the index, we verify our proposals by reproducing the already known $SU(2)$ equivariant Verlinde algebra from the Coulomb branch indices of class $\CS$ theories on the Lens space. We will see that after an appropriate normalization, the TQFT algebras on both sides are exactly identical, and so are the partition functions. In section \ref{sec:SU3}, we will use the proposed relation \eqref{Statement} to derive the $SU(3)$ equivariant Verlinde algebra from the index of $T[\Sigma,SU(3)]$ computed via the Argyres-Seiberg duality. Careful analysis of the results reveals interesting geometry of the Hitchin moduli space $\CM_H(\Sigma, SU(3))$.

\section{Equivariant Verlinde algebra and Coulomb branch index}\label{sec:EVACBI}

One obvious difference between the two brane systems \eqref{3d3d1} and \eqref{IndGeo} is that the $S^1$ factor appears on different sides of the correspondence. From the geometry of \eqref{3d3d1}, one would expect that
\be\label{Statement2}
{\genfrac{}{}{0pt}{}{\text{Equivariant Verlinde formula}}{\text{ at level $k$ on $\Sigma$}}} \quad = \quad {\genfrac{}{}{0pt}{}{\text{Partition function of}}{\text{$T[\Sigma\times S^1]$ on $L(k,1)$}}}\,.
\ee
In particular, there should be no dependence on the size of the $S^1$, so it is more natural to use ``3d variables'':
\be\label{4d3dConvert}
t=e^{L\beta-(b+b^{-1})L/r},\quad p=e^{-bL/r},\quad q=e^{-b^{-1}L/r}.
\ee
Here, $L$ is the size of the $S^1$ circle, $b$ is the squashing parameter of $L(k,1)_b$, $r$ measures the size of the Seifert base $S^2$, and $\beta$ parametrizes the ``canonical mass deformation'' of the 3d $\CN=4$ theory (in our case $T[\Sigma\times S^1]$) into 3d $\CN=2$. The latter is defined as follows on flat space. The 3d $\CN=4$ theory has R-symmetry $SU(2)_N\times SU(2)_R$ and we can view it as a 3d $\CN=2$ theory with the R-symmetry group being the diagonal subgroup $U(1)_{N+R}\subset U(1)_N\times U(1)_R$ with $U(1)_N$ and $U(1)_R$ being the Cartans of $SU(2)_N$ and $SU(2)_R$ respectively. The difference $U(1)_{N-R} = U(1)_N - U(1)_R$ of the original R-symmetry group is now a flavor symmetry $U(1)_\beta$ and we can weakly gauge it to introduce real masses proportional to $\beta$. It is exactly how the ``equivariant parameter'' in \cite{equivariant}, denoted by the same letter $\beta$, is defined.\footnote{More precisely, the dimensionless combination $\beta L$ is used. And from now on, we will rename $\beta_{\text{new}}=\beta_{\text{old}} L$ and $r_{\text{new}}=r_{\text{old}}/L$ to make all 3d variables dimensionless.}

In \cite{equivariant}, it was observed that much could be learned about the brane system \eqref{3d3d1} and the Hilbert space of complex Chern-Simons theory by preserving supersymmetry along the Lens space $L(k,1)$ in a different way, namely by doing partial topological twist instead of deforming the supersymmetry algebra. Geometrically, this corresponds to combining the last $\R^3$ factor in \eqref{3d3d2} with $L(k,1)$ to form $T^*L(k,1)$ regarded as a local Calabi-Yau 3-fold with $L(k,1)_b$ being a special Lagrangian submanifold:
\beq
\begin{matrix}
{\mbox{\textrm{$N$ fivebranes:}}} & \qquad & L(k,1)_b & \times & \Sigma & \times &S^1 \\
& \qquad & \cap  &  & \cap \\
{\mbox{\rm space-time:}} & \qquad & T^*L(k,1)_b & \times & T^* \Sigma & \times & S^1 \\
& \qquad &  \!\!\!\!\!\!\!\!\!\!\!\!\!\!\!\!\!\!\!\!\!\!\!\circlearrowright &  &\!\!\!\!\!\!\! \circlearrowright   \\
{\mbox{\rm symmetries:}} & \qquad & \!\!\!\!\!\!\!\!\!\!\!\!\!\! U(1)_R  &   & \!\!\!\! U(1)_N&.
\end{matrix}
\label{3d3dTwist}
\eeq
In this geometry, $U(1)_N$ acts by rotating the cotangent fiber of $\Sigma$, while $U(1)_R$ rotates the cotangent fiber of the Seifert base $S^2$ of the Lens space.\footnote{Note, $U(1)_N$ is always an isometry of the system whereas the $U(1)_R$ is only an isometry in certain limits where the metric on $L(k,1)$ is singular ({\it e.g.}~when $L(k,1)$ is viewed a small torus fibered over a long interval). However, if we are only interested in questions that have no dependence on the metric on $L(k,1)$, we can always assume the $U(1)_R$ symmetry to exist. For example, the theory $T[L(k,1)]$, or in general $T[M_3]$ for any Seifert manifolds $M_3$ should enjoy an extra flavor symmetry $U(1)_\beta=U(1)_N-U(1)_R$.} This point of view enables one to derive the equivariant Verlinde formula as it is now the partition function of the {\it supersymmetric} theory $T[L(k,1),\beta]$ on $\Sigma\times S^1$.

Although the geometric setting \eqref{3d3dTwist} appears to be different from the original one \eqref{3d3d1}, there is substantial evidence that they are related. For example, the equivariant Verlinde formula can be defined and computed on both sides and they agree. Namely, the partition function in the twisted background \eqref{3d3dTwist} is given by the partition function of $T[L(k,1)]$ on $\Sigma$, while the partition function under the background \eqref{3d3d1} is given by an equivariant integral over the Hitchin moduli space, and they are proven to be equal in \cite{Andersen:2016hoj}. Moreover, the modern viewpoint on supersymmetry in curved backgrounds is that the deformed supersymmetry is an extension of topological twisting, see \eg~\cite{Closset:2014uda}. Therefore, one should expect that the equivariant Verlinde formula at level $k$ could be identified with a particular slice of the four-parameter family of 4d indices $(k,p,q,t)$ (or in 3d variables $(k,\beta,b,r)$). And this particular slice should have the property that the index has no dependence on the geometry of $L(k,1)_b$. Since $T[L(k,1)]$ is derived in the limit where $L(k,1)$ shrinks, one should naturally take the $r\rightarrow 0$ limit for the superconformal index. In terms of the 4d parameters, that corresponds to
\be\label{CBI0}
p,q,t\rightarrow 0.
\ee
This is known as the Coulomb branch limit. In this particular limit, the only combination of $(k,p,q,t)$ independent of $b$ and $r$ that one could possibly construct is
\be\label{CBI1}
\ft=\frac{pq}{t}=e^{-\beta},
\ee
and this is precisely the parameter used in the Coulomb branch index. Therefore, one arrives at the following proposal:
\be\label{StatementUN}
\boxed{\genfrac{}{}{0pt}{}{\text{Equivariant Verlinde formula}}{\text{of $U(N)_k$ on $\Sigma$}}} \quad = \quad \boxed{\genfrac{}{}{0pt}{}{\text{Coulomb branch index}}{\text{of $T[\Sigma,U(N)]$ on $L(k,1)\times S^1$}}}\ .
\ee
This relation should be more accurately viewed as the natural isomorphism between two TQFT functors
\be
Z_{\text{EV}}=Z_{\text{CB}}.
\ee
At the level of partition function on a closed Riemann surface $\Sigma$, it is the equality between the equivariant Verlinde formula and the Coulomb index of $T[\Sigma]$
\be
Z_{\text{EV}}(\Sigma)=Z_{\text{CB}}(\Sigma).
\ee
Going one dimension lower, we also have an isomorphism between the Hilbert spaces of the two TQFTs on a circle:
\be
\CH_{\text{EV}}=Z_{\text{EV}}(S^1)=\CH_{\text{CB}}=Z_{\text{CB}}(S^1).
\ee
As these underlying vector spaces set the stages for any interesting TQFT algebra, the equality above is the most fundamental and needs to be established first. We now show how one can canonically identify the two seemingly different Hilbert spaces $\CH_{\text{EV}}$ and $\CH_{\text{CB}}$.

\subsection{$\CH_{\text{EV}}$ vs.~$\CH_{\text{CB}}$}

In the equivariant Verlinde TQFT, operator-state correspondence tells us that states in $\CH_{\text{EV}}$ are in one-to-one correspondence with local operators. Since these local operators come from codimension-2 ``monodromy defects'' \cite{Gukov:2006jk} (see also \cite{Gang:2015wya} in the context of 3d-3d correspondence) in $T[L(k,1)]$ supported on the circle fibers of $\Sigma\times S^1$, they are labeled by
\be
\mathbf{a}=\mathrm{diag}\{a_1,a_2,a_3,\ldots,a_N\}\in \frak{u}(N)
\ee
together with a compatible choice of Levi subgroup $\frak{L}\subset U(N)$. In the equivariant Verlinde TQFT, one only needs to consider maximal defects with $\frak{L}=U(1)^N$ as they are enough to span the finite-dimensional $\CH_{\text{EV}}$. The set of continuous parameters $\mathbf{a}$ is acted upon by the affine Weyl group $W_{\text{aff}}$ and therefore can be chosen to live in the Weyl alcove:
\be
1> a_1\geq a_2\geq\ldots\geq a_N\geq 0.
\ee
In the presence of a Chern-Simons term at level $k$, gauge invariance imposes the following integrality condition
\be\label{Integrality}
e^{2\pi i k\,\mathbf{a}}=\mathbf{1}.
\ee
We can then define
\be
\mathbf{h}=k\mathbf{a}
\ee
whose elements are now integers in the range $[0,k)$. The condition \eqref{Integrality} is also the condition for the adjoint orbit 
\be
\CO_{\mathbf{h}}=\{ghg^{-1}|g\in U(N)\}
\ee
to be quantizable. Via the Borel-Weil-Bott theorem, quantizing $\CO_{\mathbf{h}}$ gives a representation of $U(N)$ labeled by a Young tableau $\vec{h}=(h_1,h_2,\ldots,h_N)$. So, we can also label the states in $\CH_{\text{EV}}(S^1)$ by representations of $U(N)$ or, more precisely, integrable representations of the loop group of $U(N)$ at level $k$. In other words, the Hilbert space of the equivariant Verlinde TQFT is the same as that of the usual Verlinde TQFT (better known as the $G/G$ gauged WZW model). This is, of course, what one expects as the Verlinde algebra corresponds to the $\ft=0$ limit of the equivariant Verlinde algebra, and the effect of $\ft$ is to modify the algebra structure without changing $\CH_{\text{EV}}$. In particular, the dimension of $\CH_{\text{EV}}$ is independent of the value of $\ft$.

One could also use the local operators from the dimensional reduction of Wilson loops as the basis for $\CH_{\text{EV}}(S^1)$. In pure Chern-Simons theory, the monodromy defects are the same as Wilson loops. In $T[L(k,1),\beta]$ with $\beta$ turned on, these two types of defects are still linearly related by a transformation matrix, which is no longer diagonal. One of the many reasons that we prefer the maximal monodromy defects is because, under the correspondence, they are mapped to more familiar objects on the Coulomb index side. To see this, we first notice that the following brane system
\beq
\begin{matrix}
{\mbox{\textrm{ $N$ fivebranes:}}}& L(k,1)_b & \times & \Sigma & \times &S^1 \\
&   &  & \cap \\
{\mbox{\rm space-time:}}& L(k,1)_b & \times & T^* \Sigma & \times & S^1 & \times & \R^3\\
&   &  & \cup \\
{\mbox{\textrm{$n\times N$ ``defect'' fivebranes:}}} & L(k,1)_b &\times& T^*|_{p_i}\Sigma & \times & S^1
\end{matrix}
\label{3d3dDefects}
\eeq
gives $n$ maximal monodromy defects at $(p_1,p_2,\ldots,p_n)\in\Sigma$. If one first compactifies the brane system above on $\Sigma$, one obtains the 4d $\CN=2$ class $\CS$ theory $T[\Sigma_{g,n}]$ on $L(k,1)_b\times S^1$. This theory has flavor symmetry $U(N)^n$ and one can consider sectors of the theory with non-trivial flavor holonomies $\{\exp[\mathbf{a}_i],i=1,2,\ldots,n\}$ of $U(N)^n$ along the Hopf fiber. The $L(k,1)$-Coulomb branch index of $T[\Sigma_{g,n}]$ depends only on $\{\mathbf{a}_i,i=1,2,\ldots,n\}$ and therefore states in the Hilbert space $\CH_{\text{CB}}$ of the Coulomb branch index TQFT associated to a puncture on $\Sigma$ are labeled by a $U(N)$ holonomy $\mathbf{a}$. (Notice that, for other types of indices, the states are in general also labeled by a continuous parameter corresponding to the holonomy along the $S^1$ circle and the 2d TQFT for them is in general infinite-dimensional). As the Hopf fiber is the generator of $\pi_1(L(k,1))=\Z_k$, one has
\be\label{Integrality2}
e^{2\pi i k\mathbf{a}}=\mathrm{Id}.
\ee
This is exactly the same as the condition \eqref{Integrality}. In fact, we have even used the same letter $\mathbf{a}$ in both equations, anticipating the connection between the two. What we have found is the canonical way of identifying the two sets of basis vectors in the two Hilbert spaces
\be\label{StatementStates}
\begin{matrix}
\CH_{\text{EV}}^{\otimes n}& & & &\CH_{\text{CB}}^{\otimes n}\\
\rotatebox[origin=c]{90}{$\in$}& & & &\rotatebox[origin=c]{90}{$\in$}\\
\boxed{\genfrac{}{}{0pt}{}{\text{Monodromy defects on $\Sigma_{g,n}\times S^1$}}{\text{in $GL(N,\C)_k$ complex Chern-Simons theory}}}& \quad &=& \quad &\boxed{\genfrac{}{}{0pt}{}{\text{Flavor holonomy sectors}}{\text{of $T[\Sigma_{g,n}\times S^1,U(N)]$ on $L(k,1)$}}}\end{matrix}\ .
\ee
And, of course, this relation is expected as both sides are labeled by flat connections of the Chan-Paton bundle associated to the coincident $N$ ``defect'' M5-branes in \eqref{3d3dDefects}. Using the relation \eqref{StatementStates}, henceforth we identify $\CH_{\text{EV}}$ and $\CH_{\text{CB}}$. 

\subsection{The statement for a general group}

The proposed relation \eqref{Statement} between the $U(N)$ equivariant Verlinde formula and the Coulomb branch index for $T[\Sigma,U(N)]$ can be generalized to other groups. First, one could consider decoupling the center of mass degree of freedom for all coincident stacks of M5-branes. However, there are at least two different ways of achieving this. Namely, one could get rid of the $\frak{u}(1)$ part of $\mathbf{a}$ by either
\begin{enumerate}
	\item subtracting the trace part from $\mathbf{a}$:
	\be
	\mathbf{a}_{\text{SU}}=\mathbf{a}-\frac{1}{N}\mathrm{tr}\, \mathbf{a},
	\ee
	\item or forcing $\mathbf{a}$ to be traceless by imposing
	\be
	a_N=-\sum_i^{N-1}a_i
	\ee
	to get
	\be
	\mathbf{a}_{\text{PSU}}=\mathrm{diag}(a_1,a_2,\ldots,a_{N-1},-\sum_i^{N-1}a_i).
	\ee
\end{enumerate}
Naively, one may expect the two different approaches to be equivalent. However, as we are considering Lens space index, the global structure of the group comes into play. Indeed, the integrality condition \eqref{Integrality} becomes different:
\be\label{IntegralPSU}
e^{2\pi i k\cdot \mathbf{a}_{\text{SU}}}\in \Z_N=\CZ(SU(N))
\ee
while
\be\label{IntegralSU}
e^{2\pi i k\cdot \mathbf{a}_{\text{PSU}}}=\mathbf{1}=\CZ(PSU(N)).
\ee
Here $PSU(N)=SU(N)/\Z_N$ has trivial center but a non-trivial fundamental group. As a consequence of having different integrality conditions, one can get either Verlinde formula for $SU(N)$ or $PSU(N)$. In the first case, the claim is
\be\label{StatementSU}
\boxed{\genfrac{}{}{0pt}{}{\text{Equivariant Verlinde formula}}{\text{of $SU(N)_k$ on $\Sigma$}}} \quad = \quad \boxed{\genfrac{}{}{0pt}{}{\text{Coulomb branch index}}{\text{of $T[\Sigma,PSU(N)]$ on $L(k,1)\times S^1$ }}}\ .
\ee
The meaning of $T[\Sigma,PSU(N)]$ and the way to compute its Coulomb branch index will be discussed shortly. On the other hand, if one employs the second method to decouple the $U(1)$ factor, one finds a similar relation with the role of $SU(N)$ and $PSU(N)$ reversed:
\be\label{StatementPSU}
\boxed{\genfrac{}{}{0pt}{}{\text{Equivariant Verlinde formula}}{\text{of $PSU(N)_k$ on $\Sigma$}}} \quad = \quad \boxed{\genfrac{}{}{0pt}{}{\text{Coulomb branch index}}{\text{of $T[\Sigma,SU(N)]$ on $L(k,1)\times S^1$}}}\ .
\ee
Before deriving these statements, we first remark that they are all compatible with \eqref{Statement} for general $G$, 
 which we record again below:
\be\label{StatementG}
\boxed{\genfrac{}{}{0pt}{}{\text{Equivariant Verlinde formula}}{\text{of $G_k$ on $\Sigma$}}} \quad = \quad \boxed{\genfrac{}{}{0pt}{}{\text{Coulomb branch index}}{\text{of $T[\Sigma,^L\! G]$ on $L(k,1)\times S^1$}}}\ ,
\ee
since $^LU(N)=U(N)$ and $^LSU(N)=PSU(N)$. This general proposal also gives a geometric/physical interpretation of the Coulomb index of $T[\Sigma,G]$ on $L(k,1)$ by relating it to the quantization of the Hitchin moduli space $\CM_H(\Sigma,^L\!\! G)$. In fact, one can make a even more general conjecture for all 4d $\CN=2$ superconformal theories (not necessarily of class $\CS$):
\be\label{StatementTheory}
\boxed{\genfrac{}{}{0pt}{}{\text{$L(k,1)$ Coulomb index}}{\text{of a 4d $\CN=2$ superconformal theory $\CT$}}} \quad \overset{?}{=} \quad \boxed{\genfrac{}{}{0pt}{}{\text{Graded dimension of Hilbert space}}{\text{from quantization of $(\tilde{\CM}_{\CT},k\omega_I)$}}}\ .
\ee
Here, $\tilde{\CM}_{\CT}$ is the SYZ mirror \cite{Strominger:1996it} of the Coulomb branch $\CM_{\CT}$ of $\CT$ on $\R^3\times S^1$.
Indeed, $\CM_{\CT}$ has the structure of a torus fibration:
\be
\begin{matrix}
\mathbf{T}^{2d} & \hookrightarrow & \CM_{\CT} \\
  & & \downarrow \\
& & \CB \end{matrix}.
\ee
Here $\CB$ is the $d$-(complex-)dimensional Coulomb branch of $\CT$ on $\R^4$, $\mathbf{T}^{2d}$ is the 2d-torus parametrized by the holomonies of the low energy $U(1)^d$ gauge group along the spatial circle $S^1$ and the expectation values of $d$ dual photons. One can perform T-duality on $\mathbf{T}^{2d}$ to obtain the mirror manifold\footnote{In many cases, the mirror manifold $\tilde{\CM}_{\CT}=\CM_{\CT'}$ is also the 3d Coulomb branch of a theory $\CT'$ obtained by replacing the gauge group of $\CT$ with its Langlands dual. One can easily see that $\CT'$ obtained this way always has same 4d Coulomb branch $B$ as $\CT$.} $\tilde{\CM}_{\CT}$
\be
\begin{matrix}
\tilde{\mathbf{T}}^{2d} & \hookrightarrow & \tilde{\CM}_{\CT} \\
  & & \downarrow \\
& & \CB \end{matrix}.
\ee
The dual torus $\tilde{\mathbf{T}}^{2d}$ is a K\"ahler manifold equipped with a K\"ahler form $\omega$, which extends to $\omega_I$, one of the three K\"ahler forms $(\omega_I,\omega_J,\omega_K)$ of the hyper-K\"ahler manifold $\tilde{\CM}_{\CT}$. Part of the R-symmetry that corresponds to the $U(1)_N-U(1)_R$ subgroup inside the $SU(2)_R\times U(1)_N$ R-symmetry group of $\CT$ becomes a $U(1)_{\beta}$ symmetry of $\tilde{\CM}_{\CT}$.

Quantizing $\tilde{\CM}_{\CT}$ with respect to the symplectic form $k\omega_I$ yields a Hilbert space $\CH(\CT,k)$. Because $\tilde{\CM}_{\CT}$ is non-compact, the resulting Hilbert space $\CH(\CT,k)$ is infinite-dimensional. However, because the fixed point set of $U(1)_\beta$ is compact and is contained in the nilpotent cone (= the fiber of $\tilde{\CM}_{\CT}$ at the origin of $\CB$), the following graded dimension is free of any divergences and can be computed with the help of the equivariant index theorem
\be\label{CBIQuantum}
\dim_\beta\CH(\CT,k)=\sum_{m=0}^\infty \ft^m \dim \CH^{m}(\CT,k)=\int_{\tilde{\CM}_{\CT}}\mathrm{ch}(\CL^{\otimes k},\beta)\^\mathrm{Td}(\tilde{\CM}_{\CT},\beta).
\ee
Here $\ft=e^{-\beta}$ is identified with the parameter of the Coulomb branch index, $\CL$ is a line bundle whose curvature is $\omega_I$, and $\CH^{m}(\CT,k)$ is the weight-$m$ component of $\CH(\CT,k)$ with respect to the $U(1)_\beta$ action. In obtaining \eqref{CBIQuantum}, we have used the identification $\CH(\CT,k)=H^*(\tilde{\CM}_{\CT},\CL^{\otimes k})$ from geometric quantization.\footnote{One expects the higher cohomology groups to vanish, since $\CL$ is ample on each generic fiber $\tilde{\mathbf{T}}^{2d}$. For Hitchin moduli space, the vanishing of higher cohomology for $\CL^{\otimes k}$ is proven in \cite{2016arXiv160801754H,Andersen:2016hoj}.}

Now let us give a heuristic argument for why \eqref{CBIQuantum} computes the Coulomb branch index.
The Lens space $L(k,1)$ can be viewed as a torus fibered over an interval. Following \cite{Gukov:2008ve, Gukov:2010sw, Nekrasov:2010ka} and \cite{Dimofte:2011jd}, one can identify the Coulomb branch index with the partition function of a topological A-model living on a strip, with $\CM_{\CT}$ as the target space. The boundary condition at each end of the strip gives a certain brane in $\CM_{\CT}$. One can then apply mirror symmetry and turn the system into a B-model with $\tilde{\CM}_{\CT}$ as the target space. Inside $\tilde{\CM}_{\CT}$, there are two branes $\frak{B}_1$ and $\frak{B}_2$ specifying the boundary conditions at the two endpoints of the spatial interval. The partition function for this B-model computes the dimension of the $\mathrm{Hom}$-space between the two branes:
\be
Z_{\text{B-model}}=\dim \mathrm{Hom}(\frak{B}_1,\frak{B}_2).
\ee
Now $\frak{B}_1$ and $\frak{B}_2$ are objects in the derived category of coherent sheaves on $\tilde{\CM_{\CT}}$ and the quantity above can be computed using the index theorem. The equivariant version is 
\be
Z_{\text{B-model},\beta}=\dim_\beta \mathrm{Hom}(\frak{B}_1,\frak{B}_2)=\int_{\tilde{\CM}_{\CT}}\mathrm{ch}(\frak{B}_1^*,\beta)\^\mathrm{ch}(\frak{B}_2,\beta)\^\mathrm{Td}(\tilde{\CM}_{\CT},\beta). 
\ee
We can choose the duality frame such that $\frak{B}_1=\CO$ is the structure sheaf. Then $\frak{B}_2$ is obtained by acting $T^k\in SL(2,\Z)$ on $\frak{B}_1$. A simple calculation shows $\frak{B}_2=\CL^{\otimes k}$. So the Coulomb branch index indeed equals \eqref{CBIQuantum}, confirming the proposed relation \eqref{StatementTheory} (see also \cite{Fredrickson:2017yka} for a test of this relation for many Argyres-Douglas theories).

\subsubsection{$SU(N)$ vs.~$PSU(N)$}

Now let us explain why \eqref{StatementSU} and \eqref{StatementPSU} are expected. Both orbits, $\CO_{\mathbf{a_{\text{SU}}}}$ and $\CO_{\mathbf{a_{\text{PSU}}}}$, are quantizable and give rise to representations of $\frak{su}(N)$. However, as the integrality conditions are different, there is a crucial difference between the two classes of representations that one can obtain from $\mathbf{a}_{\text{SU}}$ and $\mathbf{a}_{\text{PSU}}$. Namely, one can get all representations of $SU(N)_k$ from $\CO_{\mathbf{a_{\text{SU}}}}$ but only representations\footnote{In our conventions, representations of $PSU(N)_k$ are those representations of $SU(N)_k$ invariant under the action of the center. There exist different conventions in the literature and one is related to ours by $k'=\lfloor k/N\rfloor$. Strictly speaking, when $N \nmid k$, the 3d Chern-Simons theory is not invariant under large gauge transformation and doesn't exist. Nonetheless, the 2d equivariant Verlinde algebra is still well defined and matches the algebra from the Coulomb index side.} of $PSU(N)_k$ from $\CO_{\mathbf{a_{\text{PSU}}}}$. This can be directly verified as follows.

For either $\mathbf{a}_{\text{SU}}$ or $\mathbf{a}_{\text{PSU}}$, quantizing $\CO_{\mathbf{a}}$ gives a representation of $SU(N)$ with the highest weight\footnote{Sometimes it is more convenient to use a different convention for the highest weight
\be
\vec{\lambda}=(h_1-h_2,h_2-h_3,\ldots,h_{N-1}-h_N)\equiv k\cdot (a_1-a_2,a_2-a_3,\ldots,a_{N-1}-a_N)\pmod N.
\ee
}
\be
\vec{\mu}=(h_1-h_N,h_2-h_N,\ldots,h_{N-1}-h_N)\equiv k(a_1-a_N,a_2-a_N,\ldots,a_{N-1}-a_N) \pmod N.
\label{repMu}
\ee
The corresponding Young tableau consists of $N-1$ rows with $h_i-h_N$ boxes in the $i$-th row. The integrality condition \eqref{IntegralPSU} simply says that $\vec{\mu}$ is integral. With no other constraints imposed, one can get all representations of $SU(N)$ from $\mathbf{a}_{\text{SU}}$. On the other hand, the condition \eqref{IntegralSU} requires the total number of boxes to be a multiple of $N$,
\be
\sum_{i=1}^{N-1} \mu_i=N\cdot\sum_{i=1}^{N-1}a_i \equiv 0 \pmod N,
\ee
restricting us to these representations of $SU(N)$ where the center $\Z_N$ acts trivially. These are precisely the representations of $PSU(N)$.

What we have seen is that in the first way of decoupling $U(1)$, one arrives at the equivariant Verlinde algebra for $SU(N)_k$, while the second option leads to the $PSU(N)_k$ algebra. Then, what happens on the Lens space side?

\subsubsection{$T[\Sigma,SU(N)]$ vs.~$T[\Sigma,PSU(N)]$ }\label{fluxargument}

In the second approach of removing the center, the flavor $U(N)$-bundles become well-defined $SU(N)$-bundles on $L(k,1)$ and decoupling all the central $U(1)$'s on the Lens space side simply means computing the Lens space Coulomb branch index of $T[\Sigma,SU(N)]$. So we arrive at the equivalence \eqref{StatementPSU} between $PSU(N)_k$ equivariant Verlinde algebra and the algebra of the Coulomb index TQFT for $SU(N)$. On the other hand, in the first way of decoupling the $U(1)$, the integrality condition
\be
e^{2\pi i k\cdot \mathbf{a}}=1
\ee
is not satisfied for $\mathbf{a}_{\text{SU}}$. And as in \eqref{IntegralPSU}, the right-hand side can be an arbitrary element in the center $\Z_N$ of $SU(N)$. In other words, after using the first method of decoupling the central $U(1)$, the $U(N)$-bundle over $L(k,1)$ becomes a $PSU(N)=SU(N)/\Z_N$-bundle. Another way to see this is by noticing that for $\exp[2\pi i\mathbf{a}]\in\CZ(SU(N))$,
\be
\mathbf{a}_{\text{SU}}=\mathbf{a}-\frac{1}{N}\mathrm{tr}\, \mathbf{a}= 0.
\ee
This tells us that the $U(1)$ quotient done in this way has collapsed the $\Z_N$ center of $U(N)$, giving us not a well-defined $SU(N)$-bundle but a $PSU(N)$-bundle. Therefore, it is very natural to give the name ``$T[\Sigma,PSU(N)]$'' to the resulting theory living on $L(k,1)\times S^1$, as the class $\CS$ theory $T[\Sigma,G]$ doesn't currently have proper definition in the literature if $G$ is not simply-connected.

For a general group $G$, one natural definition of the path integral of $T[\Sigma,G]$ on $L(k,1)\times S^1$ is as the path integral of $T[\Sigma,\tilde{G}]$ with summation over all possible 't Hooft fluxes labeled by $\pi_1(G)\subset\CZ(\tilde{G})$ along $L(k,1)$, where $\tilde{G}$ is the universal cover of $G$ (see \eg~\cite[Section 4.1]{Witten:2009at} for nice explanation from the 6d viewpoint). This amounts to summing over different topological types of $G$-bundles over $L(k,1)$, classified by $H^2(L(k,1),\pi_1(G))=\pi_1(G)\otimes\Z_k$.

Although this is a valid definition, it is not the right one for \eqref{Statement} to work for general $k$. This is clear from the quantization condition \eqref{IntegralPSU}, which tells us that, in order to get the $SU(N)$ Verlinde algebra, the Lens index of $T[\Sigma,PSU(N)]$ should be interpreted in the following way: in the process of assembling $\Sigma$ from pairs of pants and cylinders, we should sum over 't Hooft fluxes in the \emph{full} fundamental group $\pi_1(PSU(N))=\Z_N$, as opposed to $\Z_N\otimes\Z_k$, in the $T[\Sigma,SU(N)]$ theory for each gauge group associated with a cylinder.  But in general, $\Z_N\otimes\Z_k$ is only a proper subgroup of $\Z_N$, unless $N$ divides $k$.

However, general flux backgrounds can be realized by inserting surface operators (which we will refer to as ``flux tubes'') with central monodromy whose Levi subgroup is the entire group \cite{Gukov:2006jk}. In the spatial directions, the flux tube lives on a $S^1\subset L(k,1)$ that has linking number 1 with the Hopf fiber. So we can choose this $S^1$ to be a particular Hopf fiber $S^1_{\text{Hopf}}$. The amount of flux is labeled by an element in $\pi_1(G)\subset\CZ(\tilde{G})$. Geometrically, this construction amounts to removing a single Hopf fiber from $L(k,1)$, leading to compactly supported cohomology $H^2_c(L(k,1)\backslash S^1_{\text{Hopf}},\Z)=\Z$ that is freely generated. Then $H^2_c\left(L(k,1)\backslash S^1_{\text{Hopf}},\pi_1(G)\right)=\pi_1(G)$, and the flux can take value on the whole $\pi_1(G)$.

When $G$ is a group of adjoint type (\textit{i.e.}~$\CZ(G)$ is trivial), we will call the index of $T[\Sigma,G]$ defined this way the ``full Coulomb branch index'' of $T[\Sigma, \tilde{G}]$, which sums over \textit{all} elements of $\pi_1(G)=\CZ(\tilde{G})$. As it contains the most information about the field theory, it is also the most interesting in the whole family associated to the Lie algebra $\frak{g}$. This is not at all surprising as on the other side of the duality, the $\tilde{G}$ equivariant Verlinde algebra involves all representations of $\frak{g}$ and is the most interesting one among its cousins. 

As for the $A_{N-1}$ series that we will focus on in the rest of this paper, we will be studying the correspondence \eqref{StatementSU} between the $SU(N)$ equivariant Verlinde algebra and the Coulomb index of $T[\Sigma, PSU(N)]$. But before going any further, we will first address a common concern that the reader may have. Namely, charge quantization appears to be violated in the presence of these non-integral $SU(N)$ holonomies. Shouldn't this suggest that the index is just zero with a non-trivial flux background? Indeed, for a state transforming under the fundamental representation of $SU(N)$, translation along the Hopf fiber of $L(k,1)$ $k$ times gives a non-abelian Aharonov-Bohm phase
\be\label{ABPhase}
e^{2\pi i k \mathbf{a}_{\text{SU}}}.
\ee
Since the loop is trivial in $\pi_1(L(k,1))$, one would expect this phase to be trivial. However, in the presence of a non-trivial 't Hooft flux, \eqref{ABPhase} is a non-trivial element in the center of $SU(N)$. Then the partition function with insertion of such an 't Hooft operator is automatically zero. However, this is actually what one must have in order to recover even the usual Verlinde formula in the $\ft=0$ limit. As we will explain next, what is observed above in the $SU(2)$ case is basically the ``selection rule'' saying that in the decomposition of a tensor product
\be
\text{(half integer spin)}\otimes\text{(integer spin)}\otimes\ldots\otimes\text{(integer spin)}
\ee
there is no representation with integer spins! What we will do next is to use Dirac quantization conditions in $T[\Sigma, PSU(N)]$ to derive the selection rule above and analogous rules for the $SU(N)$ Verlinde algebra.

\subsection{Verlinde algebra and Dirac quantization}

The Verlinde formula associates to a pair of pants a fusion coefficient $f_{abc}$ which tells us how to decompose a tensor product of representations:
\be
R_a\otimes R_b=\bigoplus_{c}f_{ab}^{\phantom{ab}c}R_c.
\ee
Equivalently, this coefficient gives the dimension of the invariant subspace of three-fold tensor products
\be
\dim\mathrm{Inv}(R_a\otimes R_b\otimes R_c)=f_{abc}.
\ee
Here, upper and lower indices are related by the ``metric''
\be
\eta_{ab}=\dim\mathrm{Inv}(R_a\otimes R_b)=\delta_{a\bar{b}},
\ee
which is what the TQFT associates to a cylinder.

In the case of $SU(N)$, the fusion coefficients $f_{abc}$ are zero whenever a selection rule is not satisfied. For three representations labeled by the highest weights $\vec{\mu}^{(1)},\vec{\mu}^{(2)},\vec{\mu}^{(3)}$ in \eqref{repMu} the selection rule is
\be\label{SelectionSUN}
\sum_{i=1}^{N-1}(\mu^{(1)}_i+\mu^{(2)}_i+\mu^{(3)}_i)\equiv 0 \pmod N.
\ee
This is equivalent to the condition that $\Z_N$ acts trivially on $R_a\otimes R_b\otimes R_c$. Of course, when this action is non-trivial, it is easy to see that there can't be any invariant subspace.

Our job now is to reproduce this rule on the Coulomb index side via Dirac quantization. We start with the familiar case of $SU(2)$. The theory $T_2=T[\Sigma_{0,3},SU(2)]$ consists of eight 4d $\CN=2$ half-hypermultiplets transforming in the tri-fundamental of the $SU(2)_a\times SU(2)_b\times SU(2)_c$ flavor symmetry. The holonomy $(H_a,H_b,H_c)\in U(1)^3$ of this flavor symmetry along the Hopf fiber is given by a triple$(m_a,m_b,m_c)$ with
\be
H_I=e^{2\pi i m_I/k}, \quad I=a,b,c.
\ee
The Dirac quantization requires that the Aharonov-Bohm phase associated with a trivial loop must be trivial. So, in the presence of the non-trivial holonomy along the Hopf fiber, a physical state with charge $(e_a,e_b,e_c)$ needs to satisfy
\be
H_a^{ke_a}H_b^{ke_b}H_c^{ke_c}=e^{2\pi i \sum_{I=a,b,c} e_I m_I}=1,
\ee
or, equivalently,
\be
\sum_{I=a,b,c} e_I m_I \in \Z.
\ee
When decomposed into representations of $U(1)^3$, the tri-fundamental hypermultiplet splits into eight components:
\be
(\mathbf{2},\mathbf{2},\mathbf{2})\rightarrow \bigoplus_{\text{All $\pm$}}(\pm{ 1},\pm {1},\pm {1}).
\ee
Therefore, one needs to satisfy eight equations
\be
\pm m_a \pm m_b \pm m_c \in \Z.
\ee
For individual $m_I$, the condition is
\be\label{IntSU2}
m_I\in \frac{\Z}{2},
\ee
which is the same as the relaxed integrality condition \eqref{IntegralPSU} for $SU(2)$. This already suggests that the condition \eqref{IntegralPSU} is the most general one and there is no need to relax it further. Indeed, $m_i$ is the ``spin'' of the corresponding $SU(2)$ representation and we know that all allowed values for it are integers and half-integers.

Besides the individual constraint \eqref{IntSU2}, there is an additional one:
\be
m_a+m_b+m_c\in \Z,
\ee
which is precisely the ``selection rule'' we mentioned before. Only when this rule is satisfied, could $R_{m_c}$ appear in the decomposition of $R_{m_a}\otimes R_{m_b}$.

We then proceed to the case of $SU(N)$. When $N=3$ the theory $T_3$ doesn't have a Lagrangian description but is conjectured to have $E_6$ global symmetry \cite{Minahan:1996fg}. And the matter fields transform in the 78-dimensional adjoint representation of $E_6$ \cite{Gaiotto:2008nz, Gadde:2010te,Argyres:2007cn} which decomposes into $SU(3)^3$ representations as follows
\be
\mathbf{78}=(\mathbf{3},\mathbf{3},\mathbf{3})\oplus(\bar{\mathbf{3}},\bar{\mathbf{3}},\bar{\mathbf{3}})\oplus(\mathbf{8},\mathbf{1},\mathbf{1})\oplus(\mathbf{1},\mathbf{8},\mathbf{1})\oplus(\mathbf{1},\mathbf{1},\mathbf{8}).
\ee
The $\mathbf{8}$ is the adjoint representation of $\frak{su}(3)$ and, being a representation for both $SU(3)$ and $PSU(3)$, imposes no additional restriction on 't Hooft fluxes. So we only need to understand the quantization condition in the presence of a tri-fundamental matter $(\mathbf{3},\mathbf{3},\mathbf{3})$. A natural question, then, is whether it happens more generally, {\it i.e.}, 
\be\label{QuantSUN}
\genfrac{}{}{0pt}{}{\text{Dirac quantization condition}}{\text{ for the $T_N$ theory}} \quad = \quad \genfrac{}{}{0pt}{}{\text{Dirac quantization condition}}{\text{for a tri-fundamental matter.}}
\ee
This imposes on the $T_N$ theory an interesting condition, which is expected to be true as it turns out to give the correct selection rule for $SU(N)$ Verlinde algebra.  

Now, we proceed to determine the quantization condition for the tri-fundamental of $SU(N)^3$. We assume the holonomy in $SU(N)^3$ to be
\be
(H_a,H_b,H_c),
\ee
where
\be
H_I=\exp\left[\frac{2\pi i}{k}\mathrm{diag}\{m_{I1},m_{I2},\ldots,m_{IN}\}\right].
\ee
The tracelessness condition looks like
\be\label{Traceless}
\sum_{j=1}^N m_{Ij}=0 \quad\text{for all $I=a,b,c.$}
\ee
We now have $N^3$ constraints given by
\be
m_{aj_1}+m_{bj_2}+m_{cj_3}\in \Z \quad\text{for all choices of $j_1,j_2$ and $j_3$}.
\ee
Using \eqref{Traceless}, one can derive the individual constraint for each $i=a,b,c$:\footnote{In this paper, bold letters like $\mathbf{m}$ are used to denote an element in the Cartan subalgebra of $\frak{g}$. They are sometimes viewed as a diagonal matrix and sometimes a multi-component vector. The interpretation should be clear from the context.}
\be\label{Indiv}
\mathbf{m}_I \equiv \left(\frac{1}{N},\frac{1}{N},\frac{1}{N},\ldots,\frac{1}{N}\right)\cdot\Z \pmod \Z.
\ee
This is exactly the same as \eqref{IntegralPSU}. There is only one additional ``selection rule'' that needs to be satisfied:
\be\label{Selec}
\sum_{I=a,b,c}\sum_{j=1}^{N-1} (m_{Ij}-{m_{IN}})\equiv 0 \pmod N,
\ee
which coincides with \eqref{SelectionSUN}. Therefore, we have demonstrated the equivalence between the Dirac quantization condition of the tri-fundamental and the selection rules in the $SU(N)$ Verlinde algebra. Since the argument is independent of the value of $\ft$, the same set of selection rules also applies to the equivariant Verlinde algebra.

Beside pairs of pants, one needs one more ingredient to build a 2d TQFT---the cylinder. It can be used to glue punctures together to build general Riemann surfaces. Each cylinder corresponds to a free 4d $\CN=2$ vector multiplet. Since all of its components transform under the adjoint representation, it does not alter the individual constraints \eqref{Indiv}. However, the holonomies associated with the two punctures need to be the inverse of each other as the two flavor symmetries are identified and gauged. So the index of $T[\Sigma_{0,2},SU(N)]$ gives a diagonal ``metric''
\be
\eta_{ab}\sim \delta_{a\bar{b}}.
\ee
The proportionality constant is $\ft$ dependent and will be determined in later sections.

We can also derive the the Dirac quantization condition for $T[\Sigma_{g,n},PSU(N)]$. We use $m_{Ij}$ to label the $j$-th component of the $U(1)^N$ holonomy associated to the $I$-th puncture. Then the index or any kind of partition function of $T[\Sigma_{g,n},SU(N)]$ is zero unless
\begin{enumerate}
	\item each $\vec{m}_{I}$ satisfies the individual constraint \eqref{Indiv}, and
	\item an additional constraint analogous to \eqref{Selec},
	\be\label{SelecGeneral}
	\sum_{I=1}^n\sum_{j=1}^{N-1} (m_{Ij}-{m_{IN}})\equiv 0 \pmod N,
	\ee
	is also satisfied.
\end{enumerate}

To end this section, we will explain how the additional numerical factor in \eqref{EVFk=1} in the introduction arises from non-trivial 't Hooft fluxes. For $G=SU(N)$, one has
\be
Z_{\text{EV}}(\Sigma,k=1,\ft)=N^g\cdot\left[\frac{1}{\prod_{i=1}^{\mathrm{rank}\,G}(1-\ft^{i+1})^{2i+1}}\right]^{g-1}.
\ee
Here we are only concerned with the first factor $N^g$ which is the $k=1$ Verlinde formula for $SU(N)$
\be
Z_{\text{EV}}(\Sigma,k=1,\ft=0)=N^g.
\ee
We now derive this result on the index side.

Consider the twice-punctured torus, obtained by gluing two pairs of pants. Let $(a_1,a_2,a_3)$ and $(b_1,b_2,b_3)\in\Z_N^3$ label the 't Hooft fluxes corresponding to all six punctures. We glue $a_2$ with $b_2$, $a_3$ with $b_3$ to get $\Sigma_{1,2}$. Then we have the following set of constraints:
\be
a_2 b_2=1,\;a_3 b_3=1,
\ee
and
\be
a_1a_2a_3=1,\; b_1b_2b_3=1.
\ee
From these constraints, we can first confirm that
\be
a_1b_1=1,
\ee
which is what the selection rule \eqref{SelecGeneral} predicts. Then there is a free parameter $a_2$ that can take arbitrary values in $\Z_N$. So in the $\ft=0$ limit, the Coulomb index TQFT associates to $\Sigma_{1,2}$
\be
Z_{\text{CB}}(\Sigma_{1,2},SU(N),\ft=0)=N\delta_{a_1,\bar{b_1}}.
\ee
We can now glue $g-1$ twice-punctured tori to get
\be
Z_{\text{CB}}(\Sigma_{g-1,2},SU(N),\ft=0)=N^{g-1}\delta_{a_1,\bar{b}_{g-1}}.
\ee
Taking trace of this gives\footnote{What we have verified is basically that the algebra of $\Z_N$ 't Hooft fluxes gives the $SU(N)$ Verlinde algebra at level $k=1$, which is isomorphic to the group algebra of $\Z_N$. Another TQFT whose Frobenius algebra is also related to the group algebra of $\Z_N$ is the 2d $\Z_N$ Dijkgraaf-Witten theory \cite{dijkgraaf1990}. However, the normalizations of the trace operator are different so the partition functions are also different.}
\be
Z_{\text{CB}}(\Sigma_{g,0},SU(N),\ft=0)=N^{g}.
\ee
Combining this with the $\ft$ dependent part of \eqref{EVFk=1}, we have proved that, for $k=1$, the equivariant Verlinde formula is the same as the full Coulomb branch index.

We will now move on to cases with more general $k$ to perform stronger checks.

\section{A check of the proposal}
\label{sec:SU2}

In this section, we perform explicit computation of the Coulomb branch index for the theory $T[\Sigma_{g,n}, PSU(2)]$ in the presence of 't Hooft fluxes (or half-integral flavor holonomies). We will see that after taking into account a proper normalization, the full Coulomb branch index nicely reproduces the known $SU(2)$ equivariant Verlinde algebra. First, we introduce the necessary ingredients of 4d ${\CN}=2$ superconformal index on $S^1 \times L(k,1)$ for a theory with a Lagrangian description.

\subsection{The Lens space index and its Coulomb branch limit}

The Lens space index of 4d ${\CN} = 2$ theories is a generalization of the ordinary superconformal index on $S^1 \times S^3$, as $S^3 = L(1,1)$ \cite{Benini:2011nc}. For $k>1$, $L(k,1)$ has a nontrivial fundamental group $\mathbb{Z}_k$, and a supersymmetric theory on $L(k,1)$ tends to have a set of degenerate vacua labeled by holonomies along the Hopf fiber. This feature renders the Lens space index a refined tool to study the BPS spectra of the superconformal theory; for instance it can distinguish between theories with gauge groups that have the same Lie algebra but different topologies (\eg~$SU(2)$ versus $SO(3)$ \cite{Razamat:2013opa}). Moreover, as it involves not only continuous fugacities but also discrete holonomies, Lens space indices of class $\CS$ theories lead to a very large family of interesting and exotic 2d TQFTs \cite{Benini:2011nc,Alday:2013rs,Razamat:2013jxa}.

The basic ingredients of the Lens space index are indices of free supermultiplets, each of which can be conveniently expressed as a integral over gauge group of the plethystic exponential of the ``single-letter index'', endowed with gauge and flavor fugacities. This procedure corresponds to constructing all possible gauge invariant multi-trace operators that are short with respect to the superconformal algebra.

In particular, for a gauge vector multiplet the single-letter index is
\be\label{singleV}
f^V(p,q,t,m,k) =\frac{1}{1-p q} \pbra{\frac{p^m}{1-p^k}+\frac{q^{k-m}}{1-q^k}}(pq+\frac{pq}{t}- 1 -t) + \delta_{m,0},
\ee
where $m$ will be related to holonomies of gauge symmetries. For a half-hypermultiplet, one has
\be\label{singleH}
f^{H/2}(p,q,t,m,k) = \frac{1}{1-p q} \pbra{\frac{p^m}{1-p^k}+\frac{q^{k-m}}{1-q^k}}(\sqrt{t} - \frac{pq}{\sqrt{t}}).
\ee
In addition, there is also a ``zero point energy'' contribution for each type of field. For a vector multiplet and a half hypermultiplet, they are given by
\begin{equation}
\begin{aligned}
I_V^0(p,q,t,{\bf m}, k) & = \prod_{\alpha \in \Delta^+} \pbra{\frac{pq}{t}}^{- [\![ \alpha({\mathbf m}) ]\!]_k + \frac{1}{k}  [\![ \alpha({\mathbf m}) ]\!]_k^2},\\[0.5em]
I_{H/2}^0(p,q,t,{\mathbf m}, {\mathbf{\tilde{m}}}, k) & = \prod_{\rho \in \mathfrak{R}} \pbra{\frac{pq}{t}}^{\frac{1}{4}\pbra{[\![ \rho({\mathbf m}, {\mathbf{ \tilde{ m}}}) ]\!]_k - \frac{1}{k}  [\![ \rho({\mathbf m},  {\mathbf{\tilde {m}}}) ]\!]_k^2}},
\end{aligned}
\end{equation}
where $\dbra{x}_k$ denotes remainder of $x$ divided by $k$. The boldface letters ${\bf m}$ and ${\mathbf{ \tilde{ m}}}$ label holonomies for, respectively, gauge symmetries and flavor symmetries\footnote{As before, the holonomies are given by $e^{2\pi i\mathbf{m}/k}$.}; they are chosen to live in the Weyl alcove and can be viewed as a collection of integers $m_1 \geq m_2 \geq \dots \geq m_r$.

Now the full index can be written as
\begin{equation}
\begin{aligned}
{\CI} = \sum_{\bf m} & I_V^0(p,q,t,{\bf m})  I_{H/2}^0(p,q,t,{\bf m},{\mathbf{\tilde{m}}}) \int \prod_i \frac{dz_i}{2\pi i z_i} \Delta(z)_{\bf m}\\[0.5em]
& \times \exp \pbra{\sum_{n=1}^{+\infty}\sum_{\alpha,\rho}\frac{1}{n} \left[ f^V(p^n,q^n,t^n, \alpha({\bf m})) \alpha(z) + f^{H/2} (p^n,q^n,t^n, \rho({\bf m},{\mathbf{\tilde{m}}})) \rho(z, F) \right]}.
\label{LensIndex}
\end{aligned}
\end{equation}
Here, to avoid clutter, we only include one vector multiplet and one half-hypermultiplet. Of course, in general one should remember to include the entire field contents of the theory. Here, $F$ stands for the continuous flavor fugacities and the $z_i$'s are the gauge fugacities; for $SU(N)$ theories one should impose the condition $z_1 z_2 \dots z_N = 1$. The additional summation in the plethystic exponential is over all the weights in the relevant representations. The integration measure is determined by ${\bf m}$:
\be
\Delta_{\bf m}(z_i) = \prod_{i,j; m_i = m_j }\left(1-\frac{z_i}{z_j}\right),
\ee
since a nonzero holonomy would break the gauge group into its stabilizer.

In this paper we are particularly interested in the Coulomb branch limit, \ie~\eqref{CBI0} and \eqref{CBI1}. From the single letter index \eqref{singleV} and \eqref{singleH} we immediately conclude that $f^{H/2} = 0$ identically, so the hypermultiplets contributes to the index only through the zero point energy. As for $f^V$, the vector multiplet gives a non-zero contribution $pq/t = \ft$ for each root $\alpha$ that has $\alpha({\bf m}) = 0$. So the zero roots (Cartan generators) always contribute, and non-zero roots can only contribute when the gauge symmetry is enhanced from $U(1)^r$, \ie~when $\mathbf{m}$ is at the boundary of the Weyl alcove. This closely resembles the behavior of the ``metric'' of the equivariant Verlinde algebra, as we will see shortly. 

More explicitly, for $SU(2)$ theory, the index of a vector multiplet in the Coulomb branch limit is
\beq
{I}_V (\ft, m, k) = \ft^{- [\![ 2m ]\!]_k + \frac{1}{k}  [\![ 2m ]\!]_k^2} \pbra{\frac{1}{1-\ft}}\pbra{\frac{1}{1+\ft}}^{\delta_{\dbra{2m},0}},
\label{SU2CBIV}
\eeq
while for tri-fundamental hypermultiplet the contribution is
\beq
{I}_{H/2}(\ft, m_1,m_2,m_3,k) = \prod_{s_i = \pm} \pbra{\ft}^{\frac{1}{4}\sum_{i=1}^3 \pbra{[\![ m_i s_i ]\!]_k - \frac{1}{k}  [\![ m_i s_i ]\!]_k^2}},
\eeq
where all holonomies take values from $\{0,1/2,1,3/2,\dots k/2\}$.

Unsurprisingly, this limit fits the name of the ``Coulomb branch index.'' Indeed, in the case of $k=1$, the index receives only contributions from the Coulomb branch operators, {\it i.e.}~a collection of ``Casimir operators'' for the theory \cite{Gadde:2011uv} (\eg~${\rm{Tr}}\phi^2, \, {\rm{Tr}}\phi^3, \, \dots ,\, {\rm{Tr}}\phi^N$ for $SU(N)$, where $\phi$ is the scalar in the ${\cal N}=2$ vector multiplet). We see here that a general Lens space index also counts the Coulomb branch operators, but the contribution from each operator is modified according to the background holonomies.

Another interesting feature of the Coulomb branch index is the complete disappearance of continuous fugacities of flavor symmetries. Punctures are now only parametrized by discrete holonomies along the Hopf fiber of $L(k,1)$. This property ensures that we will obtain a \textit{finite-dimensional} algebra.

Then, to make sure that the algebra defines a TQFT, one needs to check associativity, especially because non-integral holonomies considered here are novel and may cause subtleties. We have checked by explicit computation in $\ft$ that the structure constant and metric defined by Lens space index do satisfy associativity, confirming that the ``Coulomb branch index TQFT'' is indeed well-defined. In fact, even with all $p, q, t$ turned on, the associativity still holds order by order in the expansion in terms of fugacities.

\subsection{Equivariant Verlinde algebra from Hitchin moduli space}

As explained in greater detail in \cite{equivariant}, the equivariant Verlinde TQFT computes an equivariant integral over $\CM_H$, the moduli space of Higgs bundles. In the case of $SU(2)$, the relevant moduli spaces are simple enough and one can deduce the TQFT algebra from geometry of $\CM_H$. For example, one can obtain the fusion coefficients from $\CM_H(\Sigma_{0,3}, \alpha_1, \alpha_2, \alpha_3; SU(2))$. Here the $\alpha_i$'s are the ramification data specifying the monodromies of the gauge field \cite{Gukov:2006jk} and take discrete values in the presence of a level $k$ Chern-Simons term. Since in this case the moduli space is just a point or empty, one can directly evaluate the integral. The result is as follows. 

Define $\lambda = 2 k \alpha$ whose value is quantized to be $0, 1, \dots, k$. Let
\beq
d_0 & = \lambda_1+\lambda_2+\lambda_3 - 2k,\\
d_1 & = \lambda_1 - \lambda_2 - \lambda_3,\\
d_2 & = \lambda_2 - \lambda_3 - \lambda_1,\\
d_1 & = \lambda_3 - \lambda_1 - \lambda_2,
\eeq
and moreover
\beq
\Delta \lambda = \max(d_0, d_1, d_2, d_3),
\eeq
then 
\beq
f_{\lambda_1 \lambda_2 \lambda_3} = \begin{cases}
1\ \ \ \ & {\rm{if}}\ \lambda_1+\lambda_2+\lambda_3\ \text{is even and}\ \Delta \lambda \leq 0,\\[0.5em]
\ft^{-\Delta \lambda/2}& {\rm{if}}\ \lambda_1+\lambda_2+\lambda_3\ \text{is even and}\ \Delta \lambda > 0,\\[0.5em]
0 \ & {\rm{if}}\ \lambda_1+\lambda_2+\lambda_3\ \text{is odd}.
\end{cases}
\label{POPVerlinde}
\eeq
On the other hand, the cylinder gives the trace form (or ``metric'') of the algebra 
\beq
\eta_{\lambda_1 \lambda_2} = \{ 1-\ft^2, 1-\ft, \dots, 1-\ft, 1-\ft^2 \}.
\label{CVerlinde}
\eeq
Via cutting-and-gluing, we can compute the partition function of the TQFT on a general Riemann surface $\Sigma_{g,n}$.

\subsection{Matching two TQFTs}

So far we have introduced two TQFTs: the first one is given by equivariant integration over Hitchin moduli space ${\CM}_H$,  the second one is given by the $L(k,1)$ Coulomb branch index of the theory $T[\Sigma, PSU(2)]$. It is easy to see that the underlying vector space of the two TQFTs are the same, confirming in the $SU(2)$ case the more general result we obtained previously: 
\be
Z_{\text{EV}}(S^1)=Z_{\text{CB}}(S^1).
\ee
We can freely switch between two different descriptions of the same set of basis vectors, by either viewing them as integrable highest weight representations of $\widehat{su}(2)_k$ or $SU(2)$ holonomies along the Hopf fiber. In this section, we only use highest weights $\lambda$ as the labels for puncture data, and one can easily translate them into holonomies via $\lambda = 2m$.

Then, one needs to compare the algebraic structure of the two TQFTs and may notice that there are apparent differences. Namely, if one compares $I_V$ and $I_{H/2}$ with $\eta$ and $f$ in \eqref{POPVerlinde} and \eqref{CVerlinde}, there are additional factors coming from the zero point energy in the expressions on the index side. However, one can simply rescale states in the Hilbert space on the Coulomb index side to absorb them. 

The scaling required is
\beq
|\lambda \rangle = \ft^{\frac{1}{2} \left( [\![ \lambda ]\!]_k - \frac{1}{k}  [\![ \lambda ]\!]_k^2\right)} | \lambda \rangle'.
\eeq
This makes $I_V$ exactly the same as $\eta^{\lambda \mu}$. After rescaling, the index of the half-hypermultiplet becomes
\beq
I_{H/2}\Rightarrow f'_{\lambda_1 \lambda_2 \lambda_3} = \ft^{-\frac{1}{2} \sum_{i=1}^3 \left( [\![ \lambda_i ]\!]_k - \frac{1}{k}  [\![ \lambda_i ]\!]_k^2\right)} {I}_{H/2} (\ft, \lambda_1,\lambda_2,\lambda_3,k),
\eeq
and this is indeed identical to the fusion coefficient $f_{\lambda\mu\nu}$ of the equivariant Verlinde algebra, which we show as follows. If we define
\beq
g_0 & = m_1+m_2+m_3 = \frac{1}{2} \pbra{\lambda_1+\lambda_2+\lambda_3},\\[0.5em]
g_1 & = m_1 - m_2 -m_3 = \frac{1}{2} \pbra{\lambda_1-\lambda_2-\lambda_3},\\[0.5em]
g_2 & = m_2 - m_1 - m_3 = \frac{1}{2} \pbra{\lambda_2-\lambda_1-\lambda_3},\\[0.5em]
g_3 & = m_3 - m_1 - m_2 = \frac{1}{2} \pbra{\lambda_3-\lambda_1-\lambda_3},\\[0.5em]
\eeq
then our pair of pants can be written as
\beq
f'_{\lambda_1 \lambda_2 \lambda_3} = & \ft^{\frac{1}{2k}\pbra{\dbra{g_0}_k\dbra{-g_0}_k+\dbra{g_1}_k\dbra{-g_1}_k+\dbra{g_2}_k\dbra{-g_2}_k+\dbra{g_2}_k\dbra{-g_2}_k}}\\[0.5em]
& \times \ft^{-\frac{1}{2k} \pbra{\lambda_1(k-\lambda_1)+\lambda_2(k-\lambda_2)+\lambda_3(k-\lambda_3)}}.
\eeq
Now we can simplify the above equation further under various assumptions of each $g_i$. For instance if $0< g_0 < k$ and $g_i<0$ for $i=1,2,3$, then
\beq
f'_{\lambda_1 \lambda_2 \lambda_3} = 1.
\eeq
If on the other hand, $g_0 > k$ and $g_i<0$ for $i=1,2,3$, which means $\max(g_0 - k, g_1, g_2, g_3) = g_0 - k$, then
\beq
f'_{\lambda_1 \lambda_2 \lambda_3} = t^{g_0-k}, 
\eeq
this is precisely what we obtained by \eqref{POPVerlinde}.

Therefore, we have shown that the building blocks of the two TQFTs are the same. And by the TQFT axioms, we have proven the isomorphism of the two TQFTs. For example, they both give $\ft$-deformation of the $\widehat{su}(2)_k$ representation ring; at level $k=10$ a typical example is
\beq
|3\rangle \otimes |3 \rangle = \frac{1}{1-\ft^2} |0\rangle \oplus  \frac{1}{1-\ft} |2\rangle  \oplus  \frac{1}{1-\ft} |4\rangle  \oplus  \frac{1}{1-\ft} |6\rangle  \oplus  \frac{\ft}{1-\ft} |8\rangle \oplus  \frac{\ft^2}{1-\ft^2} |10\rangle.
\eeq
For closed Riemann surfaces, we list partition functions for several low genera and levels in table \ref{SU2Partition}. And this concludes our discussion of the $SU(2)$ case.

\setlength\extrarowheight{8pt}
\begin{table}
 \begin{adjustwidth}{-2.1cm}{}
    \begin{tabular}{ | x{1.0cm} | c | c | c | c |}
      \hline
      & $k=1$ & $k=2$ & $k=3$ & $k =4$ \\[0.5em] \hline
   $g=2$ &  $\frac{4}{(1-t^2)^3}$ & $\frac{2}{(1-t^2)^3}(5t^2+6t+5)$ &  $\frac{4}{(1-t^2)^3}(4t^3+9t^2+9t+5)$ & \begin{tabular}[c]{@{}l@{}} $ \frac{1}{\left(1-t^2\right)^3} \left(16 t^4+49 t^3\right.$\\ \ \ \ $\left.+81 t^2+75 t+35\right)$ \end{tabular}\\[0.5em] \hline
   $g =3$ & $\frac{8}{(1-t^2)^6}$ &\begin{tabular}[c]{@{}l@{}}  $ \frac{4}{\left(1-t^2\right)^6} \left(9 t^4+28 t^3\right.$\\ $\left.+54 t^2+28 t+9\right) $\end{tabular} & \begin{tabular}[c]{@{}l@{}}  $\frac{8}{\left(1-t^2\right)^6}\left(8 t^6+54 t^5+159 t^4\right.$\\ $\left.+238 t^3+183 t^2+72 t+15\right) $ \end{tabular}&\begin{tabular}[c]{@{}l@{}} $\frac{1}{\left(1-t^2\right)^6} \left(64 t^8+384 t^7+1793 t^6\right.$\\ $\left.+5250 t^5+8823 t^4+8828 t^3\right.$\\ $\left.+5407 t^2+1890 t+329 \right)$  \end{tabular} \\[0.7em] \hline
   $\forall g$ & $2\left(\frac{2}{(1-t^2)^{3}}\right)^{g-1}$ & \begin{tabular}[c]{@{}l@{}} $\pbra{\frac{2 (1-t)^{2}}{(1-t^2)^{3}}}^{g-1}$\\ $+ 2\pbra{\frac{2 (1+t)^{2}}{(1-t^2)^{3}}}^{g-1}$ \end{tabular} &\begin{tabular}[c]{@{}l@{}} $ 2 \left(\frac{5+9t+9t^2+4t^3-\sqrt{5+4t}(1+5t+t^2)}{(1-t^2)^3} \right)^{g-1} + $\\ $ 2\left(\frac{5+9t+9t^2+4t^3+\sqrt{5+4t}(1+5t+t^2)}{(1-t^2)^3} \right)^{g-1}$ \end{tabular}& \begin{tabular}[c]{@{}l@{}}  $\pbra{\frac{(3+t)(1-t)^2}{(1-t^2)^3}}^{g-1} + 2\pbra{\frac{4}{1-t^2}}^{g-1}$\\ $ +\pbra{\frac{4(3+t)(1+t)^3}{(1-t^2)^3}}^{g-1}$ \end{tabular} \\[0.8cm] \hline
   \end{tabular}
   \end{adjustwidth}
\caption{The partition function $Z_{\rm{EV}}(T[L(k,1),SU(2)] , \ft) = Z_{\rm{CB}}(T[\Sigma_g,PSU(2)] , \ft)$ for genus $g=2,3$ and level $k=1,2,3,4$.} \label{SU2Partition}
\end{table}

\section{$SU(3)$ equivariant Verlinde algebra from the Argyres-Seiberg duality}\label{sec:SU3}

In the last section, we have tested the proposal about the equivalence between the equivariant Verlinde algebra and the algebra from the Coulomb index of class $\CS$ theories. Then one would ask whether one can do more with such a correspondence and what are its applications. For example, can one use the Coulomb index as a tool to access geometric and topological information about Hitchin moduli spaces? Indeed, the study of the moduli space of Higgs bundles poses many interesting and challenging problems. In particular, doing the equivariant integral directly on ${\CM}_H$ quickly becomes unpractical when one increases the rank of the gauge group. However, our proposal states that the equivariant integral could be computed in a completely different way by looking at the superconformal index of familiar SCFTs! This is exactly what we will do in this section---we will put the correspondence to good use and probe the geometry of ${\CM}_H(\Sigma, SU(3))$ with superconformal indices. 

The natural starting point is still a pair of pants or, more precisely, a sphere with three ``maximal'' punctures (for mathematicians, three punctures with full-flag parabolic structure). The 4d theory $T[\Sigma_{0,3}, SU(3)]$ is known as the $T_3$ theory \cite{Tachikawa:2015bga}, which is first identified as an ${\CN}=2$ strongly coupled rank-$1$ SCFT with a global $E_6$ symmetry\footnote{In the following we will use the name ``$T_3$ theory" and ``$E_6$ SCFT" interchangeably.} \cite{Minahan:1996fg}. In light of the proposed correspondence, one expects that the Coulomb branch index of the $T_3$ theory equals the fusion coefficients $f_{\lambda_1\lambda_2\lambda_3}$ of the $SU(3)$ equivariant Verlinde algebra. 

\subsection{Argyres-Seiberg duality and Coulomb branch index of $T_3$ theory}

\subsubsection{A short review}
As the $T_3$ theory is an isolated SCFT, there is no Lagrangian description, and currently no method of direct computation of its index is known in the literature. However, there is a powerful duality proposed by Argyres and Seiberg \cite{Argyres:2007cn}, that relates a superconformal theory with Lagrangian description at infinite coupling to a weakly coupled gauge theory obtained by gauging an $SU(2)$ subgroup of the $E_6$ flavor symmetry of the $T_3$ SCFT.

To be more precise, one starts with an $SU(3)$ theory with six hypermultiplets (call it theory A) in the fundamental representation $3\Box \oplus 3{\bar \Box}$ of the gauge group. Unlike its $SU(2)$ counterpart, the $SU(3)$ theory has the electric-magnetic duality group $\Gamma^0(2)$, a subgroup of $SL(2, \Z)$. As a consequence, the fundamental domain of the gauge coupling $\tau$ has a cusp and the theory has an infinite coupling limit. As argued by Argyres and Seiberg through direct analysis of the Seiberg-Witten curve at strong couplings, it was shown that the theory can be naturally identified as another theory B obtained by weakly gauging the $E_6$ SCFT coupled to an additional hypermultiplet in fundamental representation of $SU(2)$. There is much evidence supporting this duality picture. For instance, the $E_6$ SCFT has a Coulomb branch operator with dimension $3$, which could be identified as the second Casimir operator ${\rm{Tr}}\phi^3$ of the dual $SU(3)$ gauge group. The $E_6$ theory has a Higgs branch of $\dim_{\mathbb{C}} {\CH} = 22$ parametrized by an operator $\mathbb{X}$ in adjoint representation of $E_6$ with Joseph relation \cite{Gaiotto:2008nz}; after gauging $SU(2)$ subgroup, two complex dimensions are removed, leaving the correct dimension of the Higgs branch for the theory A. Finally, Higgsing this $SU(2)$ leaves an $SU(6) \times U(1)$ subgroup of the maximal $E_6$ group, which is the same as the $U(6) = SU(6) \times U(1)$ flavor symmetry in the A frame.

\begin{figure*}[htbp]
\begin{adjustwidth}{-2.0cm}{}
        \centering
      \begin{subfigure}[t]{.4\textwidth}
        \includegraphics[width=8cm]{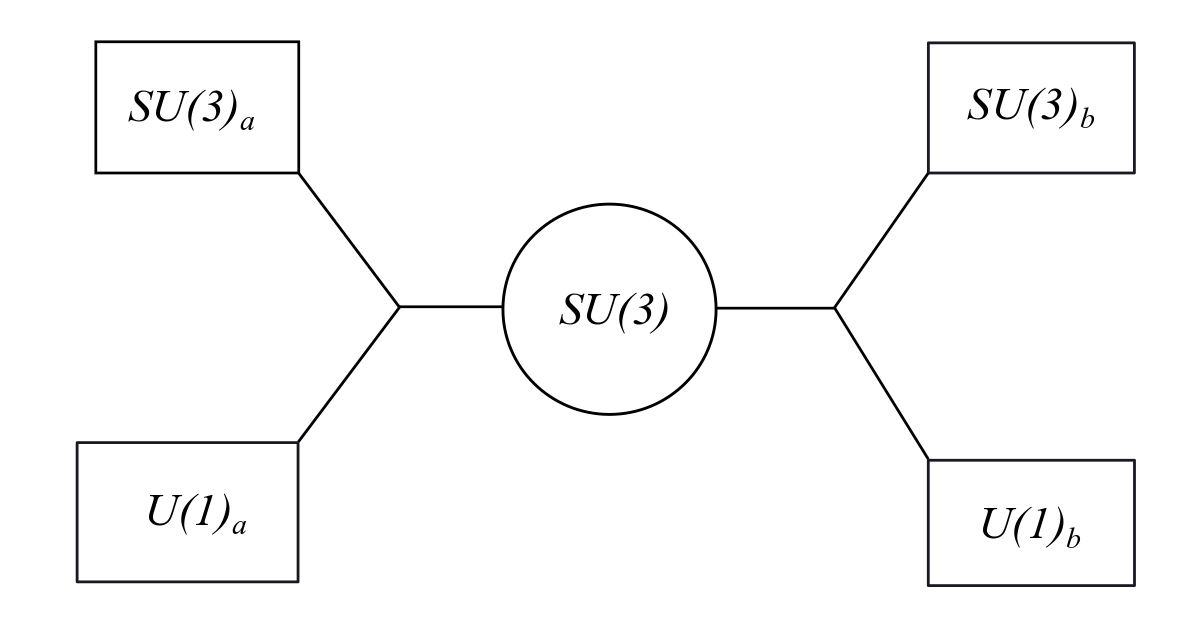}
        \caption*{(a)}
      \end{subfigure}
      \hspace{1.5cm}
      \begin{subfigure}[t]{.4\textwidth}
        \includegraphics[width=8cm]{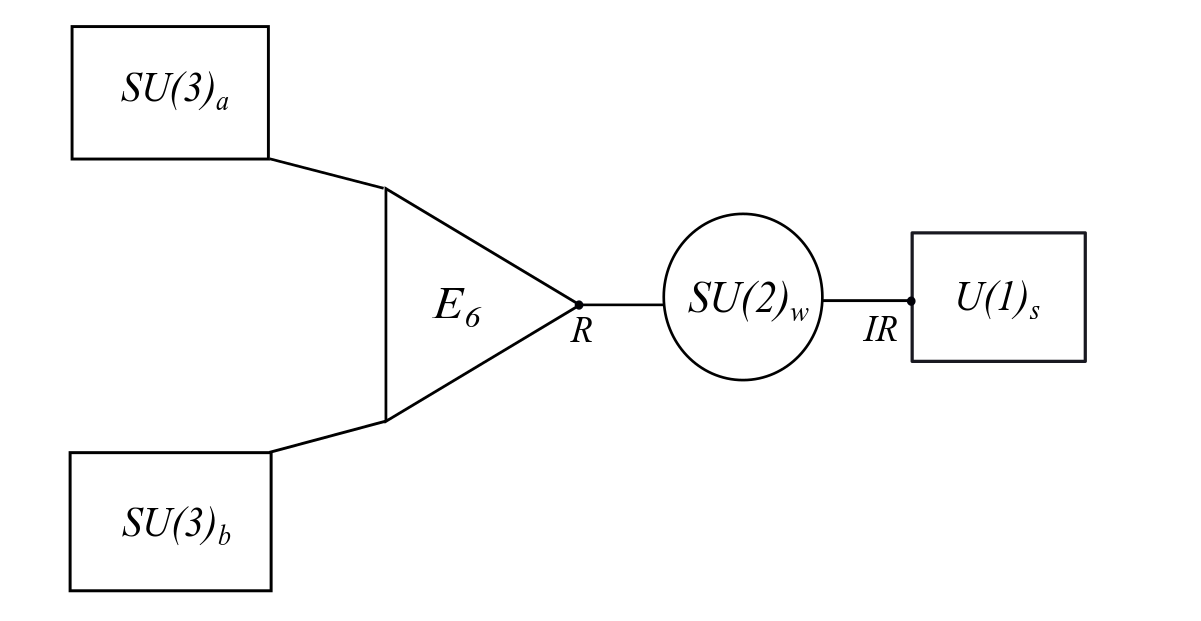}
        \caption*{(b)}
      \end{subfigure}
  \end{adjustwidth}
 \caption[Illustration of Argyres-Seiberg duality.]{Illustration of Argyres-Seiberg duality. (a) The theory A, which is an $SU(3)$ superconformal gauge theory with six hypermultiplets, with the $SU(3)_a \times U(1)_a \times SU(3)_b \times U(1)_b$ subgroup of the global $U(6)$ flavor symmetry. (b) The theory B, obtained by gauging an $SU(2)$ subgroup of the $E_6$ symmetry of $T_3$. Note in the geometric realization the cylinder connecting both sides has a regular puncture $R$ on the left and an irregular puncture $IR$ on the right.}
\label{AS3}
\end{figure*}

\begin{figure*}[htbp]
\begin{adjustwidth}{-2.0cm}{}
        \centering
      \begin{subfigure}[t]{.4\textwidth}
        \includegraphics[width=8cm]{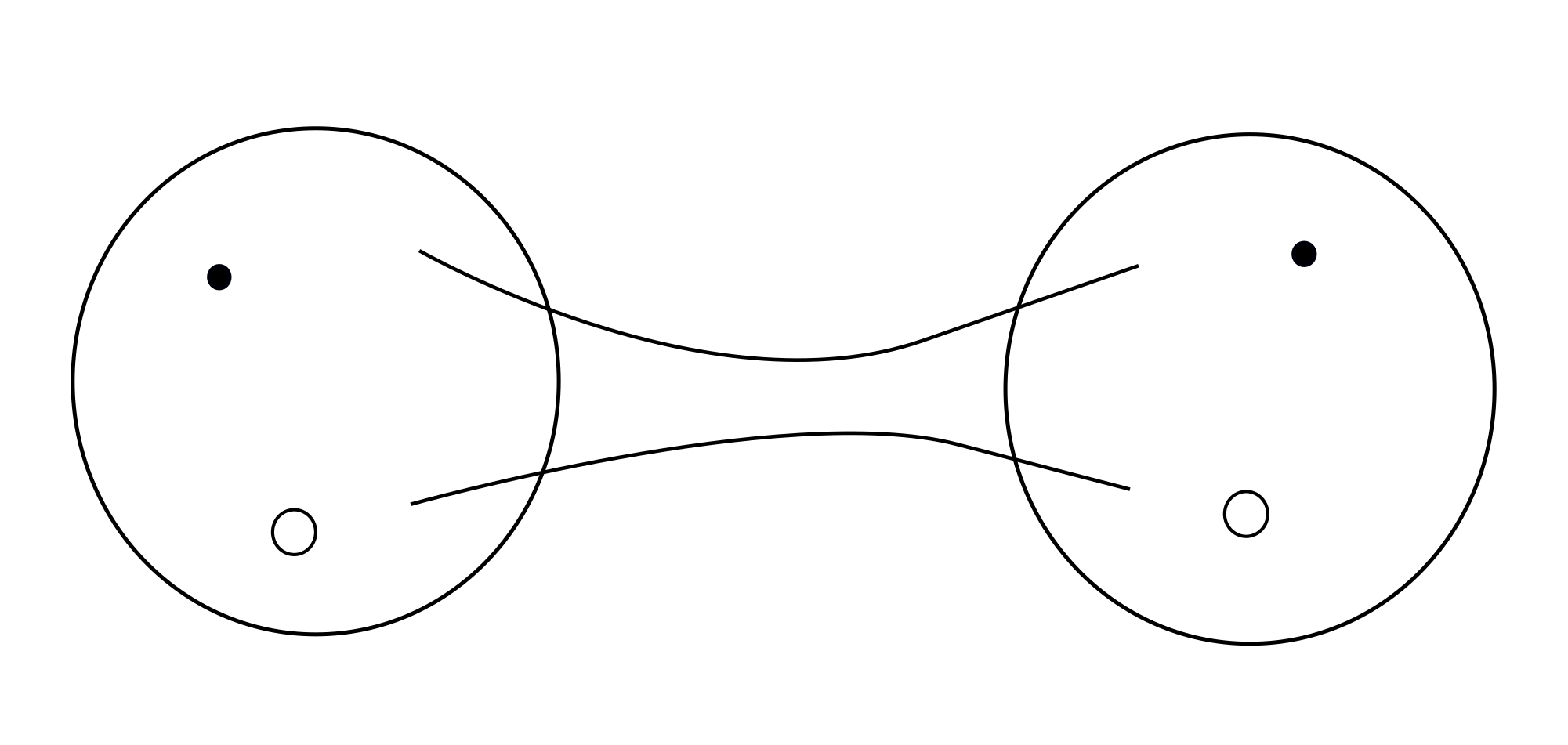}
        \caption*{(a)}
      \end{subfigure}
      \hspace{1.5cm}
      \begin{subfigure}[t]{.4\textwidth}
        \includegraphics[width=8cm]{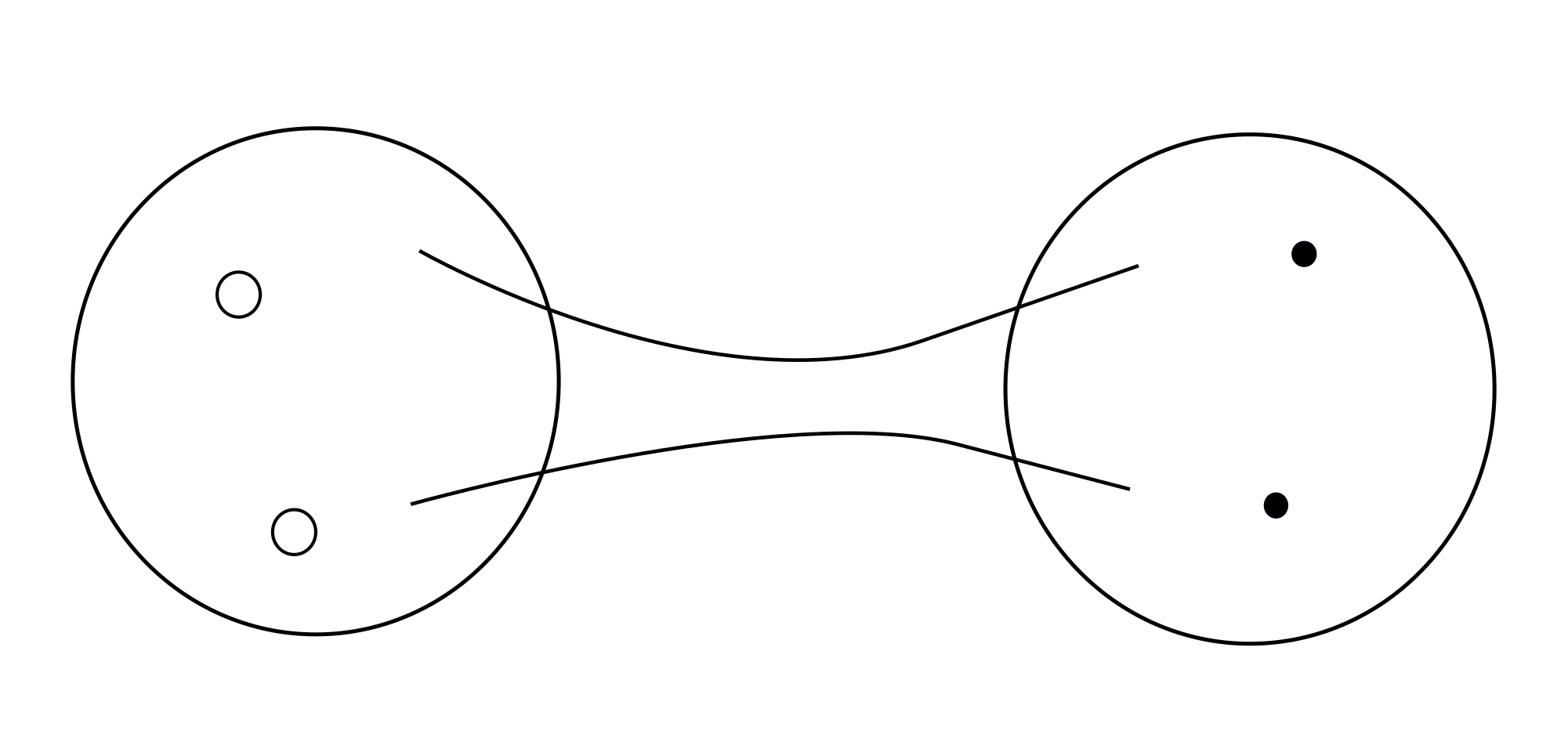}
        \caption*{(b)}
      \end{subfigure}
  \end{adjustwidth}
 \caption[Illustration of geometric realization of Argyres-Seiberg duality for $T_3$ theory.]{Illustration of geometric realization of Argyres-Seiberg duality for $T_3$ theory. The dots represent simple punctures while circles are maximal punctures. (a) The theory A, which is an $SU(3)$ superconformal gauge theory with six hypermultiplets, is pictured as two spheres connected by a long tube. Each of them has two maximal and one simple punctures. (b) The theory B, which is obtained by gauging an $SU(2)$ subgroup of the flavor symmetry of the theory $T_3$. This gauge group connects a regular puncture and an irregular puncture.}
 \label{GeneralAS1}
\end{figure*}

In \cite{Gaiotto:2009we}, the Argyres-Seiberg duality is given a nice geometric interpretation. To obtain theory A, one starts with a 2-sphere with two $SU(3)$ maximal punctures and two $U(1)$ simple punctures, corresponding to global symmetry $SU(3)_a \times SU(3)_b \times U(1)_a \times U(1)_b$, where two $U(1)$ are baryonic symmetry. In this setup, the Argyres-Seiberg duality relates different degeneration limits of this Riemann surface, see figure \ref{AS3} and \ref{GeneralAS1}.

The Argyres-Seiberg duality gives access to the superconformal index for the $E_6$ SCFT \cite{Gadde:2010te}. The basic idea is to start with the index of theory A and, with the aid of the inversion formula of elliptic beta integrals, one identifies two sets of flavor fugacities and extracts the $E_6$ SCFT index by integrating over a carefully chosen kernel. It was later realized that the above procedure has a physical interpretation, namely the $E_6$ SCFT can be obtained by flowing to the IR from an ${\CN}=1$ theory which has Lagrangian description \cite{Gadde:2015xta}. The index computation of the ${\CN}=1$ theory reproduces that of \cite{Gadde:2010te}, and the authors also compute the Coulomb branch index in the large $k$ limit.

Here we would like to obtain the index for general $k$. In principle, we could start with the ${\CN}=1$ theory described in \cite{Gadde:2015xta} and compute the Coulomb branch index on Lens space directly. However, a direct inversion is more intuitive here due to simplicity of the Coulomb branch limit, and can be generalized to arbitrary $T_N$ theories. In the next subsection we outline the general procedure of computing the Coulomb branch index of $T_3$.

\subsubsection{Computation of the index}

\begin{figure}[htbp]
\centering
\includegraphics[width=10cm]{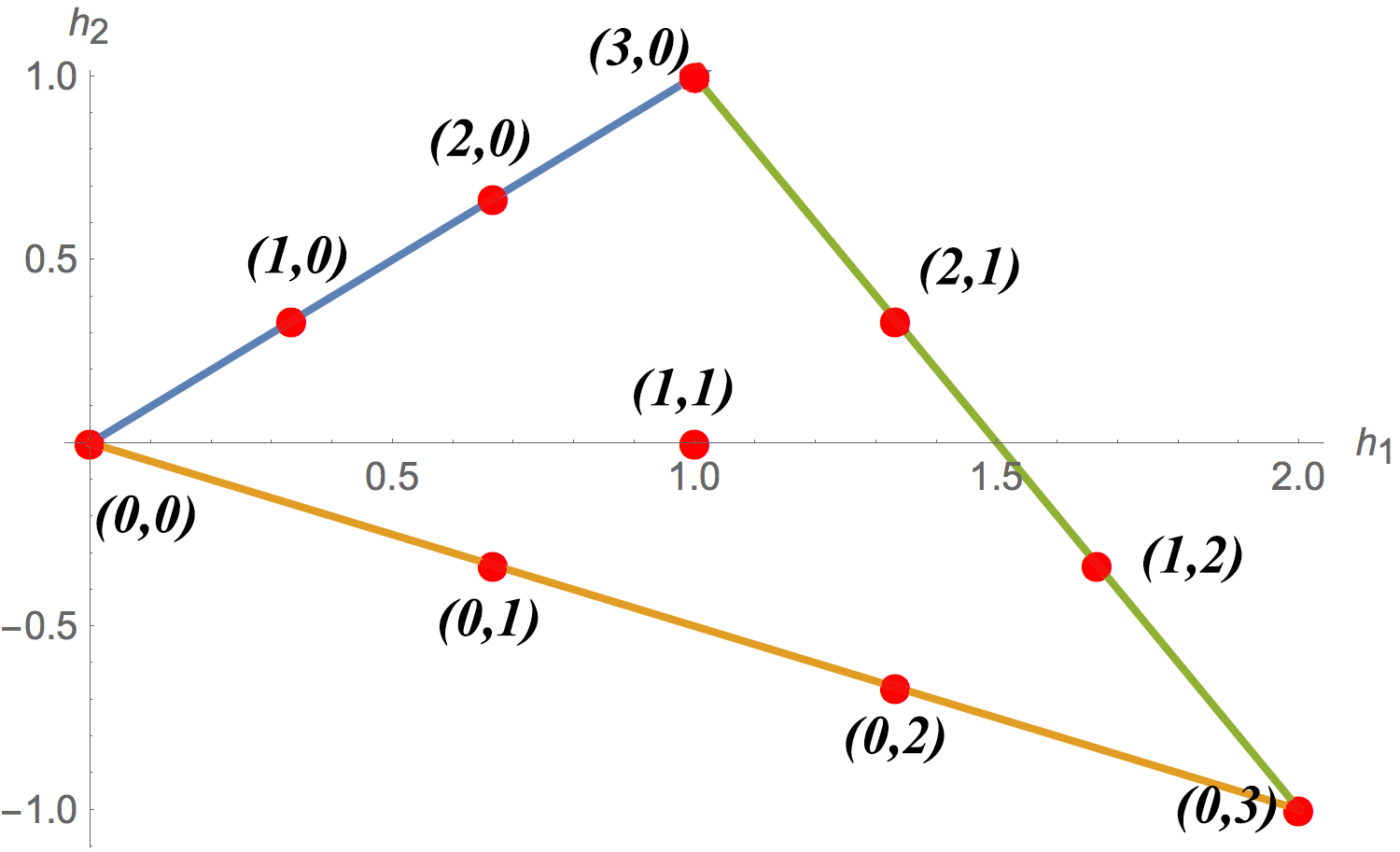}
\caption{The Weyl alcove for the choice of holonomy variables at level $k=3$. The red markers represent the allowed points. The coordinates beside each point denote the corresponding highest weight representation. The transformation between flavor holonomies and highest weight is given by \eqref{highestWeight}.}
\label{WeylAlcove}
\end{figure}

To obtain a complete basis of the TQFT Hilbert space, we need to turn on all possible flavor holonomies and determine when they correspond to a weight in the Weyl alcove. For the $T_3$ theory each puncture has $SU(3)$ flavor symmetry, so we can turn on holonomies as ${\bf h}^*=(h^*_1, h^*_2, h^*_3)$ for $* = a, b, c$ with constraints $h^*_1 + h^*_2 + h^*_3 = 0$. The Dirac quantization condition tells us that
\beq
h^r_i + h^s_j + h^t_k \in \Z
\eeq
for arbitrary $r,s,t \in \{a,b,c\}$ and $i,j,k = 1,2,3$. This means there are only three classes of choices modulo $\Z$, namely
\beq
\pbra{\frac{1}{3}, \frac{1}{3}, -\frac{2}{3}}, \ \ {\rm{or}}\ \ \pbra{\frac{2}{3}, -\frac{1}{3}, -\frac{1}{3}},\ \ {\rm{or}}\ \ \pbra{0,0,0}\ \ \pmod{\mathbb{Z}}.
\eeq
Furthermore, the three punctures either belong to the same class (for instance, all are $(1/3,1/3,-2/3) \pmod \Z$) or to three distinct classes. Recall that the range of the holonomy variables are also constrained by the level $k$, so we pick out the Weyl alcove as the following:
\beq\label{fundDomain}
D(k) = \{ (h_1,h_2, h_3) | h_1 \geq h_2, h_1 \geq -2h_2, 2h_1 + h_2 \leq k \},
\eeq
with a pictorial illustration in figure \ref{WeylAlcove}.

As we will later identify each holonomy as an integrable highest weight representation for the affine Lie algebra $\widehat{su}(3)_k$, it is more convenient to use the label $(\lambda_1, \lambda_2)$ defined as
\beq
\lambda_1 = h_2 - h_3, \ \ \ \lambda_2 = h_1 - h_2. 
\label{highestWeight}
\eeq
They are integers with $\lambda_1 + \lambda_2 \leq k$ and $(\lambda_1,\lambda_2)$ lives on the weight lattice of $su(3)$. The dimension of the representation with the highest weight $(\lambda_1, \lambda_2)$ is
\beq
\dim R_{(\lambda_1,\lambda_2)} = \frac{1}{2}(\lambda_1+1)(\lambda_2+1)(\lambda_1+\lambda_2+2).
\eeq

Next we proceed to compute the index in the Coulomb branch limit. As taking the Coulomb branch limit simplifies the index computation dramatically, one can easily write down the index for theory A\footnote{In \cite{Gadde:2015xta} the authors try to compensate for the non-integral holonomies of $n_a$ and $n_b$ by shifting the gauge holonomies ${\bf m}$. In contrast, our approach is free from such subtleties because we allow non-integral holonomies for all flavor symmetries as long as the Dirac quantization condition is obeyed.}:
\begin{equation}
\begin{aligned}
{\CI}_A& (\ft, {\mathbf{\tilde{m}}}_a, {\mathbf{\tilde{m}}}_b, n_a, n_b)  \\[0.5em]
& = \sum_{\bf m} I_{H/2} (\ft, {\bf m},{\mathbf{\tilde{m}}}_a, n_a) \int \prod_{i=1}^2 \frac{dz_i}{2\pi i  z_i} \Delta(z)_{\bf m} I_V(\ft, z, {\bf m})  I_{H/2} (\ft, -{\bf m}, {\mathbf{\tilde{m}}}_b, n_b),
\end{aligned}
\end{equation}
where ${\bf m}_a, {\bf m}_b$ and $n_a, n_b$ denote the flavor holonomies for $SU(3)_{a,b}$ and $U(1)_{a,b}$ respectively
. It is illustrative to write down what the gauge integrals look like:
\begin{equation}
I_V(\ft, {\bf m}) =\int \prod_{i=1}^2 \frac{dz_i}{2\pi i  z_i} \Delta(z)_{\bf m} I_V(\ft, z, {\bf m}) = I_V^0(\ft, {\bf m}) \times \begin{cases} \frac{1}{(1-t^2)(1-t^3)}, \ \ m_1 \equiv m_2 \equiv m_3\  \pmod k,\\[0.5em]
\frac{1}{(1-t)(1-t^2)}, \ \ m_i \equiv m_j \neq m_k\  \pmod k,\\[0.5em]
\frac{1}{(1-t)^2},\ \ m_1 \neq m_2 \neq m_3\ \pmod k.
\end{cases}
\end{equation}
Except the zero point energy $I_V^0(\ft, {\bf m})$ the rest looks very much alike our ``metric" for the $SU(3)$ equivariant Verlinde TQFT. Moreover,
\be
I_{H/2} ({\bf m}, {\mathbf{\tilde{m}}}_a, n_a)= \prod_{\psi \in R_{\Phi}} \ft^{\frac{1}{4} \pbra{ [\![ \psi({\bf m},{\mathbf{\tilde{m}}}_a,n_a) ]\!]_k - \frac{1}{k}  [\![ \psi({\bf m},{\mathbf{\tilde{m}}}_a,n_a) ]\!]_k^2}},
\ee
where for a half-hypermultiplet in the fundamental representation of $SU(3) \times SU(3)_a$ with positive $U(1)_a$ charge we have
\be
\psi_{ij}( {\bf m}, {\mathbf{\tilde{m}}}_a, n_a) = {\bf m}_i + {\mathbf{\tilde{m}}}_{a,j} + n_a.
\ee

Now we write down the index for theory B. Take the $SU(3)_a \times SU(3)_b \times SU(3)_c$ maximal subgroup of $E_6$ and gauge $SU(2)$ subgroup of the $SU(3)_c$ flavor symmetry. This leads to the replacement
\be
\{h_{c,1}, h_{c,2}, h_{c,3} \} \rightarrow \{ w + n_y, n_y - w, -2 n_y \},
\label{E6 gauge}
\ee
where $n_y$ denotes the fugacity for the remaining $U(1)_y$ symmetry, and $n_s$ is the fugacity for $U(1)_s$ flavor symmetry rotating the single hypermultiplet. We then write down the index of theory B as
\beq
{\CI}_B (\ft, {\bf h}_a, {\bf h}_b, n_y, n_s) = \sum_w C^{E_6} ({\bf h}_a, {\bf h}_b, w, n_y) I_V (\ft, w) I_{H/2} (-w, n_s),
\label{theoryBindex}
\eeq
where $I_V (\ft, w)$ is given by \eqref{SU2CBIV} with substitution $m \rightarrow w$, and $w = 0, 1/2, \dots, k/2$. Argyres-Seiberg duality tells us that
\beq
{\CI}_A& (\ft, {\mathbf{\tilde{m}}}_a, {\mathbf{\tilde m}}_b, n_a, n_b) = {\CI}_B (\ft, {\bf h}_a, {\bf h}_b, n_y, n_s),
\eeq
with the following identification of the holonomy variables:
\beq
{\mathbf{\tilde{m}}_a} & = {\bf h}_a, \ \ {\mathbf{\tilde{m}}_b} = {\bf h}_b; \\[0.5em]
n_a & = \frac{1}{3}n_s - n_y, \ \ n_b = - \frac{1}{3}n_s - n_y.
\label{ChangeHolonomy}
\eeq

On the right-hand side of the expression \eqref{theoryBindex} we can view the summation as a matrix multiplication with $w$ and $n_s$ being the row and column indices respectively. Then we can take the inverse of the matrix $I_{H/2} (-w, n_s)$, $I^{-1}_{H/2} (n_s, w')$, by restricting the range\footnote{As long as it satisfies the Dirac quantization condition, we do not have to know what the range of $n_s$ should be. For example, $n_s = 0,1/2,\dots, k/2$ is a valid choice.} of $n_s$ to be the same as $w$ and multiply it to both sides of \eqref{theoryBindex}. This moves the summation to the other side of the equation and gives:
\beq
\boxed{C^{E_6} (\ft, {\bf h}_a, {\bf h}_b, w, n_y,k) =\sum_{n_s}  \frac{1}{ I_V (\ft, w)}{\CI}_A (\ft, {\bf h}_a, {\bf h}_b, n_a, n_b,k) I_{H/2}^{-1} (n_s, w) }\ .
\eeq

We now regard $C^{E_6} (\ft, {\bf h}_a, {\bf h}_b, {\bf h}_c,k)$ as the fusion coefficient of the 2d equivariant Verlinde algebra, and have checked the associativity. Moreover, let us confirm that the index obtained in this way is symmetric under permutations of the three $SU(3)$ flavor fugacities, and the flavor symmetry group is indeed enhanced to $E_6$.  First of all, we have permutation symmetry for three $SU(3)$ factors at, for instance, level $k=2$:
\beq
C^{E_6}\pbra{\frac{2}{3}, \frac{2}{3}, 0, 0, \frac{4}{3}, -\frac{2}{3}} = C^{E_6}\pbra{\frac{2}{3}, \frac{2}{3}, \frac{4}{3}, -\frac{2}{3}, 0, 0} = \dots = C^{E_6}\pbra{ \frac{4}{3}, -\frac{2}{3},\frac{2}{3}, \frac{2}{3}, 0, 0} = \frac{1+\ft^4}{1-\ft^3}.
\eeq
To show that the index $C^{E_6}$ is invariant under the full $E_6$ symmetry, one needs to show that the two $SU(3)$ factors, combined with the $U(1)_y$ symmetry, enhance to an $SU(6)$ symmetry. The five Cartan elements of this $SU(6)$ group can be expressed as the combination of the fluxes \cite{Gadde:2015xta}:
\beq
\pbra{h_1^a - n_y, h_2^a - n_y, -h_1^a- h_2^a - n_y, h^b_1+n_y, h^b_2+n_y}.
\eeq
Then the index should be invariant under the permutation of the five Cartans. Note the computation is almost the same as in \cite{Gadde:2015xta} except that not all permutations necessarily exist---an allowed permutation should satisfy the charge quantization condition. Restraining ourselves from the illegal permutations, we have verified that the global symmetry is enlarged to $E_6$.

Finally, at large $k$ our results reproduce these of \cite{Gadde:2015xta}, as can be checked by analyzing the large $k$ limit of the matrix $ I_{H/2}^{-1} (n_s, w) $. Indeed, at large $k$ the matrix $I_{H/2}(w, n_s)$ can be simplified as
\beq
I_{H/2} = \ft^{\frac{1}{2}(|w+n_s| + |-w+n_s|)} = \left(
\begin{array}{ccccccc} 1 & 0 & \ft & 0 & \ft^2 & 0 & \dots \\
                                    0 & \sqrt{\ft} & 0 & \ft^{\frac{3}{2}} & 0 & \ft^{\frac{5}{2}} & \\
                                    \ft  &  0  & \ft & 0 &  \ft^2 & 0 & \\
                                    0 & \ft^{\frac{3}{2}} & 0 & \ft^{\frac{3}{2}} & 0 & \ft^{\frac{5}{2}} \\
                                    \ft^2 & 0 & \ft^2 & 0 & \ft^2 & 0 &\\
                                    0 & \ft^{\frac{5}{2}} & 0 & \ft^{\frac{5}{2}} & 0 & \ft^{\frac{5}{2}}\\
                                    \vdots & & & & & & \ddots
\end{array}
\right).
\eeq
Upon inversion it gives
\beq
I^{-1}_{H/2} = \left( \begin{array}{ccccccc} \frac{1}{1-\ft} & 0 & -\frac{1}{1-\ft} & 0 & 0 & 0 & \dots \\
                                    0 & \frac{1}{\sqrt{\ft}(1-\ft)} & 0 & -\frac{1}{\sqrt{\ft}(1-\ft)} & 0 & 0 & \\
                                    -\frac{1}{1-\ft}  &  0  & \frac{1+\ft}{\ft(1-\ft)} & 0 &  -\frac{1}{\ft(1-\ft)} & 0 & \\
                                    0 & -\frac{1}{\sqrt{\ft}(1-\ft)} & 0 & \frac{1+\ft}{\ft^{\frac{3}{2}}(1-\ft)} & 0 &  -\frac{1}{\ft^{\frac{3}{2}}(1-\ft)} \\
                                    0 & 0 & -\frac{1}{\ft(1-\ft)} & 0 & \frac{1+\ft}{\ft^2(1-\ft)} & 0 &\\
                                    0 & 0 & 0 & -\frac{1}{\ft^{\frac{3}{2}}(1-\ft)} & 0 & \frac{1+\ft}{\ft^{\frac{5}{2}}(1-\ft)}\\
                                    \vdots & & & & & & \ddots
\end{array} \right).
\eeq
Here $w$ goes from $0, 1/2, 1, 3/2, \cdots$. For a generic value of $w$ only three elements in a single column can contribute to the index\footnote{By ``generic" we mean the first and the second column are not reliable due to our choice of domain for $w$. It is imaginable that if we take $w$ to be a half integer from $(-\infty, +\infty)$, then such ``boundary ambiguity" can be removed. But we refrain from doing this to have weights living in the Weyl alcove.}. For large $k$ the index of vector multiplet becomes
\beq
I_V(w) = \ft^{-2w} \pbra{\frac{1}{1-\ft}},
\eeq
and we get
\beq
C^{E_6} (\ft, {\bf h}_a, {\bf h}_b, w, n_y)  = \ft^w & \left[(1+\ft){\CI}_A (\ft, {\bf h}_a, {\bf h}_b, n_y, w,k) \right.\\[0.5em]
& \left.- \ft~ {\CI}_A (\ft, {\bf h}_a, {\bf h}_b, n_y, w-1,k) - {\CI}_A (\ft, {\bf h}_a, {\bf h}_b, n_y, w+1,k)\right],
\eeq
which exactly agrees with \cite{Gadde:2015xta}.

\subsection{$SU(3)$ equivariant Verlinde algebra}

Now with all the basic building blocks of the 2d TQFT at our disposal, we assemble the pieces and see what interesting information could be extracted.

The metric of the TQFT is given by the Coulomb branch index of an $SU(3)$ vector multiplet, with a possible normalization factor. Note the conjugation of representations acts on a highest weight state $(\lambda_1,\lambda_2)$ via
\beq
\overline{ (\lambda_1,\lambda_2)} = (\lambda_2, \lambda_1),
\eeq
and the metric $\eta^{\lambda\mu}$ is non-vanishing if and only if $\mu = \bar \lambda$. Let
\beq
N(\lambda_1, \lambda_2, k) = \ft^{-\frac{1}{k}\left(\dbra{\lambda_1}_k\dbra{-\lambda_1}_k+\dbra{\lambda_2}_k\dbra{-\lambda_2}_k+\dbra{\lambda_1+\lambda_2}_k\dbra{-\lambda_1-\lambda_2}_k\right)},\\[0.5em]
\eeq
and we rescale our TQFT states as
\be
( \lambda_1, \lambda_2 )' = N(\lambda_1, \lambda_2, k)^{-\frac{1}{2}} (\lambda_1, \lambda_2).
\ee
Then the metric $\eta$ takes a simple form (here we define $\lambda_3 = \lambda_1 + \lambda_2$):
\be
\eta^{(\lambda_1, \lambda_2)\overline{(\lambda_1, \lambda_2)}} = \begin{cases} \frac{1}{(1-t^2)(1-t^3)}, \ \ \text{ if}\ \  \dbra{\lambda_1}_k = \dbra{\lambda_2}_k = 0,\\[0.3em]
\frac{1}{(1-t)(1-t^2)}, \ \ \ \text{ if only one}\ \  \dbra{\lambda_i}_k = 0\ \ \text{for}\ \ i = 1,2,3,\\[0.3em]
\frac{1}{(1-t)^2}, \ \ \ \ \ \ \ \ \ \text{ if all}\ \  \dbra{\lambda_i}_k \neq 0.
 \end{cases}
 \label{SU(3)C}
\ee

Next we find the ``pair of pants'' $f_{\pbra{\lambda_1,\lambda_2}\pbra{\mu_1,\mu_2}\pbra{\nu_1,\nu_2}}$, from the normalized Coulomb branch index of $E_6$ SCFT:
 \beq
 f_{\pbra{\lambda_1,\lambda_2}\pbra{\mu_1,\mu_2}\pbra{\nu_1,\nu_2}} = \pbra{N(\lambda_1, \lambda_2, k)N(\mu_1, \mu_2, k)N(\nu_1, \nu_2, k)}^{\frac{1}{2}}C^{E_6} (\ft, \lambda_1,\lambda_2; \mu_1,\mu_2; \nu_1,\nu_2; k).
 \label{SU(3)PoP}
 \eeq
Along with the metric we already have, they define a $\ft$-deformation of the $\widehat{su}(3)_k$ fusion algebra. For instance we could write down at level $k=3$:
\beq
(1,0) \otimes (1,0) & = \frac{1+\ft+\ft^3}{(1-\ft)(1-\ft^2)(1-\ft^3)}(0,1)\oplus\frac{1+2\ft^2}{(1-\ft) (1-\ft^2) (1-\ft^3)}(2,0)\\[0.5em]
& \oplus \frac{\ft(2+\ft)}{(1-\ft)(1-\ft^2)(1-\ft^3)}(1,2).
\eeq
Using dimensions to denote representations, the above reads
\beq
{\bf 3} \times {\bf 3}  = & \frac{1+\ft+\ft^3}{(1-\ft)(1-\ft^2)(1-\ft^3)}{\mathbf{\bar 3}}+\frac{1+2\ft^2}{(1-\ft) (1-\ft^2) (1-\ft^3)}{\bf 6}\\[0.5em]
& + \frac{\ft(2+\ft)}{(1-\ft)(1-\ft^2)(1-\ft^3)}{\mathbf{ \bar{15}}}.
\eeq
When $\ft = 0$, it reproduces the fusion rules of the affine $\widehat{su}(3)_k$ algebra, and $f_{\lambda\mu\nu}$ becomes the fusion coefficients $N_{\lambda \mu \nu}^{(k)}$. These fusion coefficients are worked out combinatorically in \cite{Gepner:1986wi, Kirillov:1992np,Begin:1992rt}. We review details of the results in appendix \ref{sec: FusionCoeff}.

With pairs of pants and cylinders, one can glue them together to get the partition function on a closed Riemann surface, which gives the $SU(3)$ equivariant Verlinde formula: a $\ft$-deformation of the $SU(3)$ Verlinde formula. For genus $g=2$, at large $k$, one can obtain
\beq
& \dim_{\beta} {\CH}_{{\rm CS}} (\Sigma_{2,0}; SL(3,{\mathbb{C}}), k)\\[0.5em]
& = \frac{1}{20160}k^8 + \frac{1}{840} k^7 + \frac{7}{480}k^6 + \frac{9}{80}k^5 + \frac{529}{960}k^4 + \frac{133}{80}k^3 + \frac{14789}{5040}k^2 + \frac{572}{210} k + 1\\[0.5em]
& + \pbra{\frac{1}{2520}k^8 + \frac{1}{84} k^7 + \frac{17}{120}k^6 + \frac{17}{20}k^5 + \frac{319}{120}k^4 + \frac{15}{4}k^3 + \frac{503}{2520}k^2 - \frac{1937}{420} k - 3}\ft\\[0.5em]
& + \pbra{\frac{1}{560}k^8 + \frac{9}{140} k^7 + \frac{31}{40}k^6 + \frac{39}{10}k^5 + \frac{727}{80}k^4 + \frac{183}{20}k^3 + \frac{369}{140}k^2 - \frac{27}{70} k + 1}\ft^2\\[0.5em]
& + \dots,
\eeq
and the reader can check that the degree zero piece in $\ft$ is the usual $SU(3)$ Verlinde formula for $g=2$ \cite{guha2009witten}:
\beq
\dim & {\CH}(\Sigma_{g,0}; SU(3), k)\\[0.5em]
& = \frac{(k+3)^{2g-2} 6^{g-1}}{2^{7g-7}} \sum_{\lambda_1, \lambda_2} \pbra{\sin \frac{\pi(\lambda_1+1)}{k+3}\sin \frac{\pi (\lambda_2+1)}{k+3}\sin \frac{\pi(\lambda_1+\lambda_2+2)}{k+3}}^{2-2g},
\eeq
expressed as a polynomial in $k$.

For a 2d TQFT, the state associated with the ``cap" contains interesting information, namely the ``cap state" tells us how to close a puncture. Moreover, there are many close cousins of the cap. There is one type which we call the ``central cap'' that has a defect with central monodromy with the Levi subgroup being the entire gauge group (there is no reduction of the gauge group when we approach the singularity). For $SU(3)$ equivariant Verlinde algebra, besides the ``identity-cap'' the central cap also includes ``$\omega$-cap'' and ``$\omega^2$-cap,'' and the corresponding TQFT states are denoted by $|\phi \rangle_1, |\phi \rangle_\omega$ and $| \phi \rangle_{\omega^2}$. One can also insert on the cap a minimal puncture (gauge group only reduces to $SU(2)\times U(1)$ as opposed to $U(1)^3$ for maximal punctures) and the corresponding states can be expressed as linear combinations of the maximal puncture states which we use as the basis vectors of the TQFT Hilbert space. 

The cap state can be deduced from $f$ and $\eta$ written in \eqref{SU(3)PoP} and \eqref{SU(3)C}, since closing a puncture on a three-punctured sphere gives a cylinder. In algebraic language,
\beq
f_{\lambda \mu \phi} = \eta_{\lambda\mu}.
\eeq
One can easily solve this equation, obtaining
\beq
| \phi \rangle_1 = |0, 0 \rangle - \ft(1+\ft) | 1,1 \rangle + \ft^2 |0,3 \rangle + \ft^2 |3,0 \rangle - \ft^3 |2,2 \rangle.
\label{1cap}
\eeq

For other two remaining caps, by multiplying\footnote{More precisely, we multiply holonomies with these central elements and translate the new holonomies back to weights.} $\omega$ and $\omega^2$ on the above equation \eqref{1cap}, we obtain 
\beq\label{CentralCap}
| \phi \rangle_{\omega} & = |k, 0 \rangle - \ft(1+\ft) | k-2,1 \rangle + \ft^2 |k-3,0 \rangle + \ft^2 |k-3,3 \rangle - \ft^3 |k-4,2 \rangle,\\[0.5em]
| \phi \rangle_{\omega^2} & = |0, k \rangle - \ft(1+\ft) | 1,k-2 \rangle + \ft^2 |0,k-3 \rangle + \ft^2 |3,k-3 \rangle - \ft^3 |2,k-4 \rangle.
\eeq
When closing a maximal puncture using $| \phi \rangle_{\omega}$, we have a ``twisted metric'' $\eta'_{\lambda \mu}$ which is non-zero if and only if $(\mu_1,\mu_2) = (\lambda_1, k-\lambda_1 - \lambda_2)$. When closing a maximal puncture using $| \phi \rangle_{\omega^2}$, we have another twisted metric $\eta''_{\lambda \mu}$ which is non-zero if and only if $(\mu_1,\mu_2) = (k-\lambda_1 - \lambda_2, \lambda_2)$. When there are insertions of central monodromies on the Riemann surface, it is easier to incorporate them into twisted metrics instead of using the expansion \eqref{CentralCap}. 

For minimal punctures, the holonomy is of the form $(u,u,-2u)$, modulo the action of the affine Weyl group, where $u$ takes value $0, 1/3, 2/3, \dots, k - 2/3, k-1/3$. We can use index computation to expand the corresponding state $|u\rangle_{U(1)}$ in terms of maximal punctures. After scaling by a normalization constant
\beq
\ft^{\frac{1}{2} \pbra{\dbra{3u}_k - \frac{1}{k}\dbra{3u}^2_k}},
\eeq
the decomposition is given by the following:
\begin{enumerate}
\item[(1).] $\langle 0, 0 \rangle - \ft^2 \langle 1,1 \rangle$, if $k = u$ or $u=0$;
\item[(2).] $\langle 3u, 0 \rangle - \ft \langle 3u-1,2 \rangle$, if $k > 3u >0$;
\item[(3).] $\langle 3u, 0 \rangle - \ft^2 \langle 3u-2,1 \rangle$, if $k = 3u$;
\item[(4).] $\langle 2k-3u, 3u-k \rangle - \ft \langle 2k-3u-1,3u-k-1 \rangle$, if $3u/2 < k < 3u$;
\item[(5).] $\langle 0, 3u/2 \rangle - \ft^2 \langle 1, 3u/2-2 \rangle$, if $k = 3u/2$;
\item[(6).] $\langle 0, 3k-3u \rangle - \ft \langle 2, 3k-3u-1 \rangle$, if $u < k < 3u/2$.
\end{enumerate}
The above formulae have a natural ${\mathbb{Z}}_2$-symmetry of the form ${\mathcal{C}} \circ \psi$, where
\beq 
\psi:  (u,k) \rightarrow (k-u,k),
\eeq
and $\CC$ is the conjugation operator that acts linearly on Hilbert space:
\beq
{\CC}:  (\lambda_1, \lambda_2) \rightarrow (\lambda_2, \lambda_1), \ \ \ (\lambda_1,\lambda_2) \in {\CH}.
\eeq
This $\mathbb{Z}_2$ action sends each state in the above list to itself. Moreover, it is interesting to observe that when $\ft = 0$, increasing $u$ from $0$ to $k$ corresponds to moving along the edges of the Weyl alcove ($c.f.$ figure \ref{WeylAlcove}) a full cycle. This may not be a surprise because closing a maximal puncture actually implies that one only considers states whose $SU(3)$ holonomy $(h_1, h_2, h_3)$ preserves at least $SU(2) \subset SU(3)$ symmetry, which are precisely the states lying on the edges of the Weyl alcove. 

\subsection{From algebra to geometry}

This TQFT structure reveals a lot of interesting geometric properties of moduli spaces of rank $3$ Higgs bundles. But as the current paper is a physics paper, we only look at a one example---but arguably the most interesting one---the moduli space ${\CM}_H(\Sigma_{0,3}, SU(3))$. In particular this moduli space was studied in \cite{Gothen1994,garcia2004betti} and \cite{boalch2012hyperkahler} from the point of view of differential equations. Here, from index computation, we can recover some of the results in the mathematical literature and reveal some new features for this moduli space. In particular, we propose the following formula for the fusion coefficient $f_{\lambda\mu\nu}$:
\be
f_{\pbra{\lambda_1,\lambda_2}\pbra{\mu_1,\mu_2}\pbra{\nu_1,\nu_2}} = \ft^{k\eta_0} \pbra{\frac{k{\rm{Vol}}(\CM) + 1}{1-\ft} + \frac{2\ft}{(1-\ft)^2} }+ \frac{Q_1(\ft)}{(1-\ft^{-1})(1-\ft^2)} + \frac{Q_2(\ft)}{(1-\ft^{-2})(1-\ft^3)}.
\label{SU(3)Equiv}
\ee
This ansatz comes from Atiyah-Bott localization of the equivariant integral done in similar fashion as in \cite{equivariant}. The localization formula enables us to write the fusion coefficient $f$ in \eqref{SU(3)PoP} as a summation over fixed points of the $U(1)_H$ Hitchin action. In \eqref{SU(3)Equiv}, $\eta_0$ is the moment map\footnote{Recall the $U(1)_H$ Hitchin action is generated by a Hamiltonian, which we call $\eta$---not to be confused with the metric, which will make no appearance from now on. $\eta$ is also the norm squared of the Higgs field.} for the lowest critical manifold $\CM$. When the undeformed fusion coefficients $N_{\lambda\mu\nu}^{(k)} \neq 0$, one has
\beq
k{\rm{Vol}}(\CM) + 1 = N_{\lambda\mu\nu}^{(k)}, \ \ \eta _0 = 0.
\eeq

Numerical computation shows that $Q_{1,2}(\ft)$ are individually a sum of three terms of the form
\beq
Q_1(\ft) = \sum_{i=1}^3 \ft^{k\eta_i}, \ \ \ \ Q_2(\ft) = \sum_{j=4}^6 \ft^{k\eta_j},
\eeq
where $\eta_i$ are interpreted as the moment maps at each of the six higher fixed points of $U(1)_H$.

\begin{figure}[htbp]
\centering
\includegraphics[width=11cm]{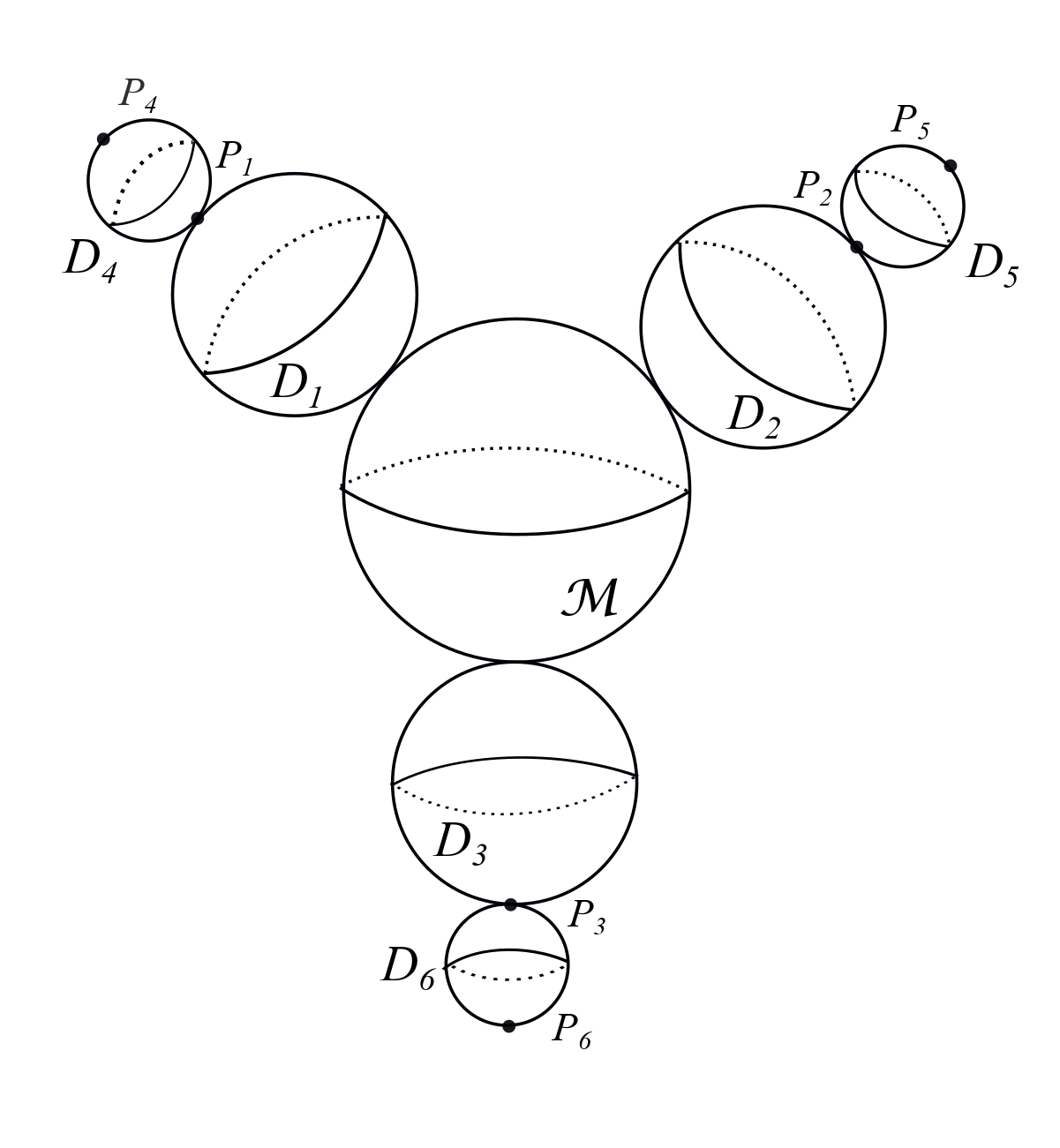}
\caption[The illustration of the nilpotent cone in ${\CM}_H(\Sigma_{0,3}, SU(3))$]{The illustration of the nilpotent cone in ${\CM}_H(\Sigma_{0,3}, SU(3))$. Here $\CM$ is the base $\cp^1$, $D_{1,2,3}$ consist of downward Morse flows from $P_{1,2,3}$ to the base, while $D_{4,5,6}$ include the flows from $P_{4,5,6}$ to $P_{1,2,3}$.}
\label{SingularFiber}
\end{figure}

The moduli space $\CM$ of $SU(3)$ flat connections on $\Sigma_{0,3}$ is either empty, a point or $\cp^1$ depending on the choice of $(\lambda,\mu,\nu)$ \cite{hayashi1999moduli}, and when it is empty, the lowest critical manifold of $\eta$ is a $\cp^1$ with $\eta_0>0$ and we will still use $\CM$ to denote it. The fixed loci of $\CM_H(\Sigma_{0,3}, SU(3))$ under $U(1)$ action consist of $\CM$ and the six additional points, and there are Morse flow lines traveling between them. The downward Morse flow coincides with the nilpotent cone \cite{Biswas}---the singular fiber of the Hitchin fibration, and its geometry is depicted in figure \ref{SingularFiber}. The Morse flow carves out six spheres that can be divided into two classes. Intersections of $D^{(1)}_i \bigcap D^{(2)}_i$ are denoted as $P^{(1)}_{1,2,3}$, and at the top of these $D^{(2)}_i$'s there are $P^{(2)}_{1,2,3}$. We also use $P_1,\ldots,P_6$ and $D_1, \ldots, D_6$ sometimes to avoid clutter. The nilpotent cone can be decomposed into
\be
{\CN} = {\CM} \cup D_i^{(1)} \cup D_j^{(2)},
\ee
which gives an affine $E_6$ singularity (IV$^*$ in Kodaira's classification) of the Hitchin fibration. Knowing the singular fiber structure, we can immediately read off the Poincar\'{e} polynomial for ${\CM}_H(\Sigma_{0,3}, SU(3))$:
\beq
\CP_{\fr} = 1 + 7 \fr^2,
\eeq
which is the same as that given in \cite{garcia2004betti}.

To use the Atiyah-Bott localization formula, we also need to understand the normal bundle to the critical manifolds. For the base, the normal bundle is the cotangent bundle with $U(1)_H$ weight 1. Its contribution to the fusion coefficient is given by
\be
t^{k\eta_0}\int_{\CM}\frac{\mathrm{Td}(\cp^1)\^e^{k\omega}}{1-e^{-\beta+2\omega'}}= \ft^{k\eta_0} \pbra{\frac{k{\rm{Vol}}(\CM) + 1}{1-\ft} + \frac{2\ft}{(1-\ft)^2}}.
\ee
For the higher fixed points, the first class $P^{(1)}$ has normal bundle $\C[-1]\oplus\C[2]$ with respect to $U(1)_H$, which gives a factor
\be
\frac{1}{(1-\ft^{-1})(1-\ft^2)}
\ee
multiplying $e^{k\eta_{1,2,3}}$. For the second class $P^{(2)}$, the normal bundle is $\C[-2]\oplus\C[3]$ and we instead have a factor
\be
\frac{1}{(1-\ft^{-2})(1-\ft^3)}.
\ee

In this paper, we won't give the analytic expression for the seven moment maps and will leave \eqref{SU(3)Equiv} as it is. Instead, we will give a relation between them:
\beq
2k & = 6 ( N^{(k)}_{\lambda\mu\nu} - 1) + 3k(\eta_1+\eta_2 + \eta_3) + k(\eta_4+\eta_5+\eta_6)\\[0.5em]
     & =  6 k {\rm{Vol}}(\CM) + 3k(\eta_1+\eta_2 + \eta_3) + k(\eta_4+\eta_5+\eta_6).
\eeq
This is verified numerically and can be explained from geometry. Noticing that the moment maps are related to the volume of the $D$'s:
\beq
{\rm{Vol}}(D_1) & = \eta_1,\ \ {\rm Vol}(D_2) = \eta_2, \ \ {\rm Vol}(D_3) = \eta_3,\\[0.5em]
{\rm{Vol}}(D_4) & =\frac{ \eta_4-\eta_1}{2},\ \ {\rm Vol}(D_5) = \frac{ \eta_5-\eta_2}{2}, \ \ {\rm{Vol}}(D_6) = \frac{ \eta_6-\eta_3}{2}.
\label{ModuliVolume}
\eeq
The factor $2$ in the second line of \eqref{ModuliVolume} is related to the fact that $U(1)_H$ rotates the $D^{(2)}$'s twice as fast as it rotates the $D^{(1)}$'s. Then we get the following relation between the volume of the components of $\CN$:
\beq
{\rm{Vol}}(\mathbf{F}) = 6 {\rm Vol}(\CM) + 4\sum_{i=1}^3 {\rm Vol}(D_i) + 2\sum_{i=4}^6{\rm Vol}(D_j).
\label{Vrelation}
\eeq
Here $F$ is a generic fiber of the Hitchin fibration and has volume
\beq
{\rm{Vol}}(\mathbf{F}) = 2.
\eeq

\begin{figure}[htbp]
\centering
\includegraphics[width=10cm]{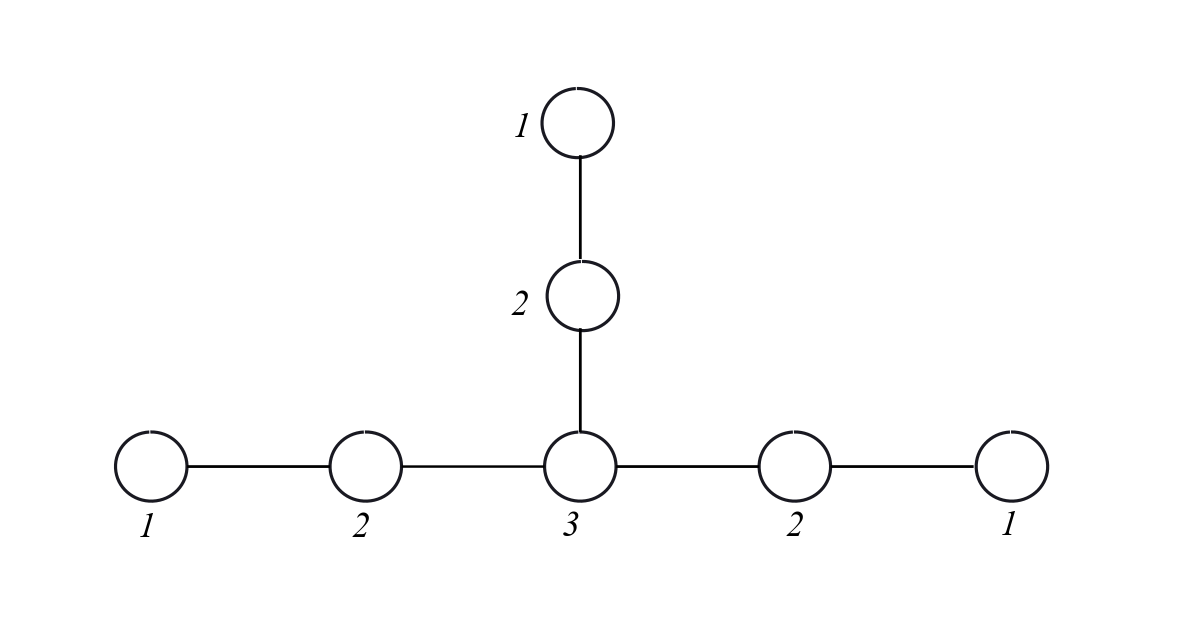}
\caption{The affine $\widehat{E}_6$ extended Dynkin diagram. The Dynkin label gives the multiplicity of each node in the decomposition of the null vector.}
\label{E6Diag}
\end{figure}

The intersection form of different components in the nilpotent cone gives the Cartan matrix of affine $E_6$. Figure \ref{E6Diag} is the Dynkin diagram of $\hat{E}_6$, and coefficients in \eqref{Vrelation} are Dynkin labels on the corresponding node. These numbers tell us the combination of $D$'s and $\CM$ that give a null vector $\mathbf{F}$ of $\hat{E}_6$.

\subsection{Comments on $T_N$ theories}

The above procedure can be generalized to arbitrary rank, for all $T_N$ theories, if we employ the generalized Argyres-Seiberg dualities. 
There are in fact several ways to generalized Argyres-Seiberg duality \cite{Tachikawa:2015bga, Gaiotto:2009we, Chacaltana:2010ks}. For our purposes, we want no punctures of the $T_N$ to be closed under dualities, so we need the following setup \cite{Gaiotto:2009we}.

\begin{figure*}[htbp]
\begin{adjustwidth}{-2.0cm}{}
        \centering
      \begin{subfigure}[t]{.4\textwidth}
        \includegraphics[width=8cm]{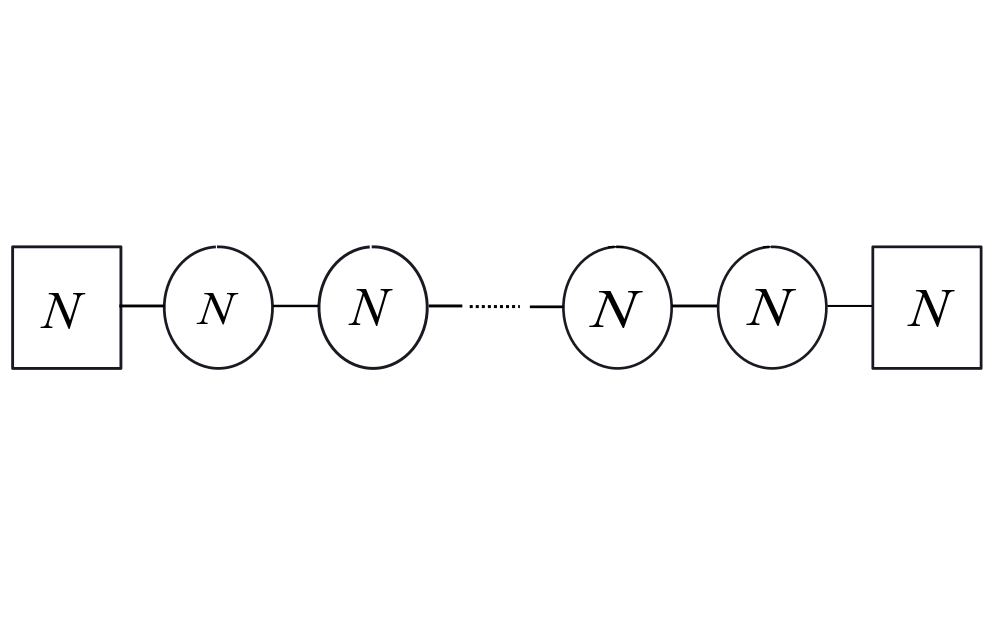}
        \caption*{(a)}
      \end{subfigure}
      \hspace{1.5cm}
      \begin{subfigure}[t]{.4\textwidth}
        \includegraphics[width=8cm]{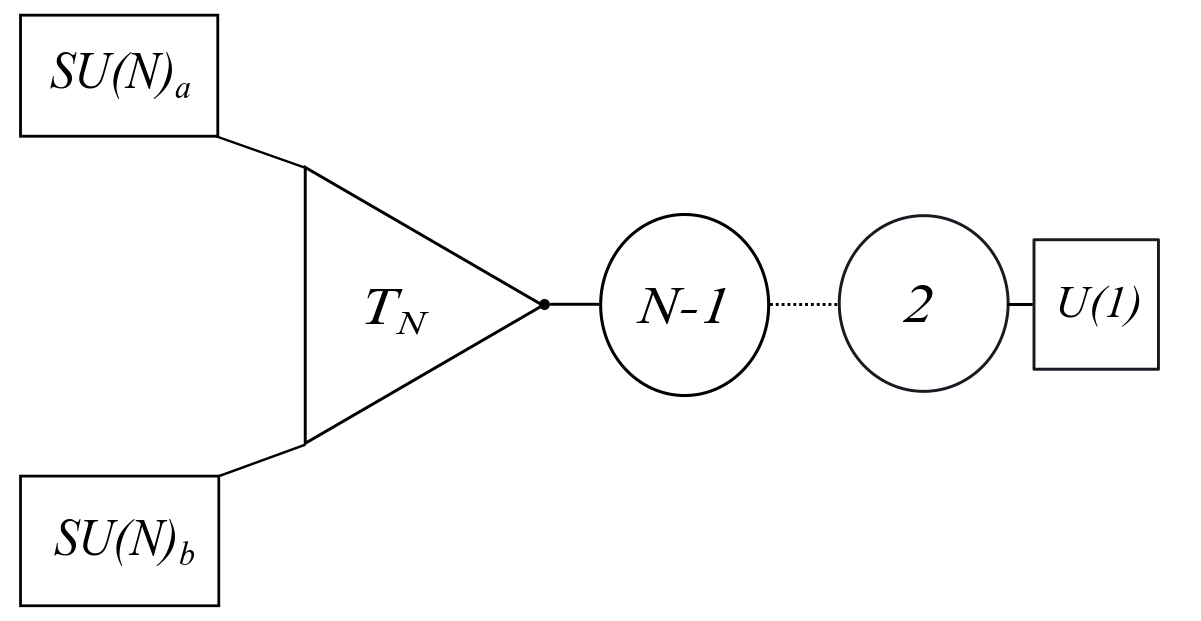}
        \caption*{(b)}
      \end{subfigure}
  \end{adjustwidth}
 \caption[Illustration of generalized Argyres-Seiberg duality for the $T_N$ theories.]{Illustration of generalized Argyres-Seiberg duality for the $T_N$ theories. (a) The theory A', which is a linear quiver gauge theory with $N-2$ $SU(N)$ vector multiplets. Between each gauge node there is a bi-fundamental hypermultiplet, and at each end of the quiver there are $N$ fundamental hypermultiplets. In the quiver diagram we omit the $U(1)^{N-1}$ baryonic symmetries. (b) The theory B' is obtained by gauging an $SU(N-1)$ subgroup of the $SU(N)^3$ flavor symmetry of $T_N$, giving rise to a quiver tail. Again the $U(1)$ symmetries are implicit in the diagram.}
\label{TN_AS}
\end{figure*}

\begin{figure*}[htbp]
\begin{adjustwidth}{-2.0cm}{}
        \centering
      \begin{subfigure}[t]{.4\textwidth}
        \includegraphics[width=9cm]{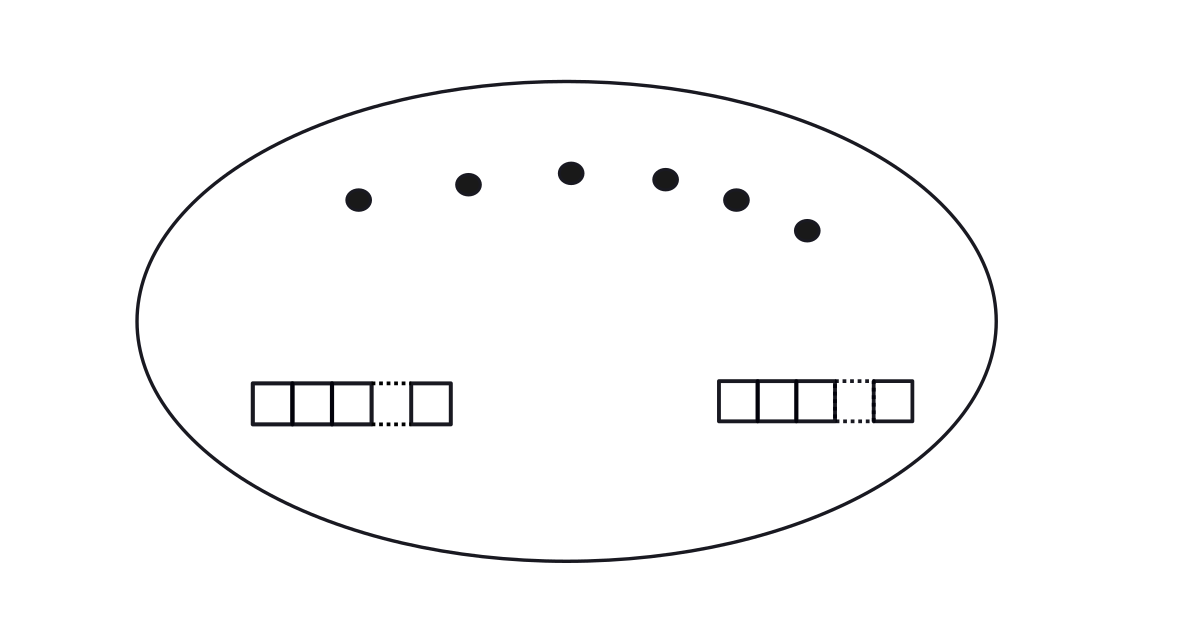}
        \caption*{(a)}
      \end{subfigure}
      \hspace{1.5cm}
      \begin{subfigure}[t]{.4\textwidth}
        \includegraphics[width=9cm]{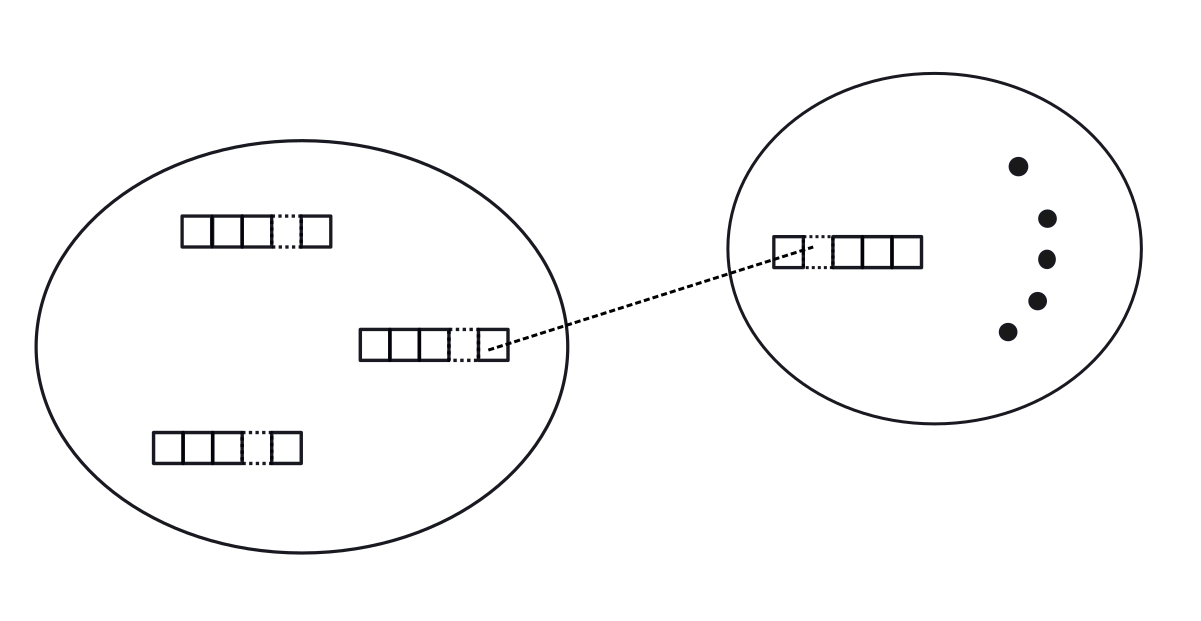}
        \caption*{(b)}
      \end{subfigure}
  \end{adjustwidth}
 \caption[Illustration of the geometric realization of generalized Argyres-Seiberg duality for $T_N$ theories.]{Illustration of the geometric realization of generalized Argyres-Seiberg duality for $T_N$ theories. (a) The theory A' is obtained by compactifying 6d $(2,0)$ theory on a Riemann sphere with two maximal $SU(N)$ punctures and $N-1$ simple punctures. (b) The theory B', obtained by colliding $N-1$ simple punctures, is then the theory that arises from gauging a $SU(N-1)$ flavor subgroup of $T_N$ by a quiver tail.}
\label{TN_ASGeo}
\end{figure*}

We start with a linear quiver gauge theory A' with $N-2$ nodes of $SU(N)$ gauge groups, and at each end of the quiver we associate $N$ hypermultiplets in the fundamental representation of $SU(N)$. One sees immediately that each gauge node is automatically superconformal. Geometrically, we actually start with a punctured Riemann sphere with two full $SU(N)$ punctures and $N-1$ simple punctures. Then, the $N-1$ simple punctures are brought together and a hidden $SU(N-1)$ gauge group becomes very weak. In our original quiver diagram, such a procedure of colliding $N-1$ simple punctures corresponds to attaching a quiver tail of the form $SU(N-1) - SU(N-2) - \cdots - SU(2)$ with a single hypermultiplet attached to the last $SU(2)$ node. See figure \ref{TN_AS} for the quiver diagrams and figure \ref{TN_ASGeo} for the geometric realization.

Here we summarize briefly how to obtain the Lens space Coulomb index of $T_N$. Let ${\CI}^N_{A'}$ be the index of the linear quiver theory, which depends on two $SU(N)$ flavor holonomies ${\bf h}_a$ and ${\bf h}_b$ (here we use the same notation as that of $SU(3)$) and $N-1$ $U(1)$-holonomies $n_i$ where $i=1,2,\dots, N-1$. In the infinite coupling limit, the dual weakly coupled theory B' emerges. One first splits the $SU(N)_c$ subgroup of the full $SU(N)^3$ flavor symmetry group into $SU(N-1)\times U(1)$ and then gauges the $SU(N-1)$ part with the first gauge node in the quiver tail. As in the $T_3$ case there is a transformation:
\beq
\pbra{h^c_1, h^c_2, \cdots, h^c_N} \rightarrow \pbra{w_1, w_2, \cdots w_{N-2}, {\tilde n}_0}.
\eeq
After the $SU(N-1)$ node, there are $N-2$ more $U(1)$ symmetries, we will call those associated holonomies ${\tilde n}_j$ with $j=1,2,\dots, N-2$. Again there exists a correspondence as in the $T_3$ case:
\beq
\pbra{n_1, n_2, \dots, n_{N-1}} \rightarrow \pbra{{\tilde n}_0, {\tilde n}_1, \dots, {\tilde n}_{N-2}}.
\eeq
Then the Coulomb branch index of the theory B' is
\beq
{\CI}^N_{B'} ({\bf h}^a, {\bf h}^b, {\tilde n}_0, {\tilde n}_1, \dots, {\tilde n}_{N-2}) = \sum_{\{w_i\}} C^{T_N}({\bf h}^a, {\bf h}^b, w_1, w_2, \cdots w_{N-2}, {\tilde n}_0) {\CI}_{T}(w_i; {\tilde n}_1, \dots, {\tilde n}_{N-2}),
\eeq
where ${\CI}_T$ is the index of the quiver tail:
\beq
{\CI}_T (w_i; {\tilde n}_1, \dots, {\tilde n}_{N-2}) = & \sum_{\{w^{(N-2)}_i\}} \sum_{\{w^{(N-3)}_i\}} \dots \sum_{\{w^{(2)}_i\}}{I}^V_{N-1} (w_i) {I}^H_{N-1,N-2} (w_i, w^{(N-2)}_j, {\tilde n}_1){I}^V_{N-2} (w^{(N-2)}_i)\\[0.5em]
& \times {I}^H_{N-2,N-3} (w^{(N-2)}_i, w^{(N-3)}_j, {\tilde n}_2) {I}^V_{N-3} (w^{(N-3)}_i)  \times \dots \\[0.5em]
& \times {I}^V_{2} (w^{(2)}_i) {I}^H_{2,1} (w^{(2)}_i, {\tilde n}_{N-2}).
\eeq

Now we can view ${\CI}_T$ as a large matrix $\mathfrak{M}_{\{w_i\}, \{{\tilde n}_j\}}$, and in fact it is a square matrix. Although the set $\{{\tilde n}_j\}$ appears to be bigger, there is an affine Weyl group $\hat{A}_{N-2}$ acting on it. From the geometric picture, one can directly see the $A_{N-2}=S_{N-2}$ permuting the $N-2$; and the shift $n_i\rightarrow n_i+k$, which gives the same holonomy in $U(1)_i$, enlarges the symmetry to that of $\hat{A}_{N-2}$. After taking quotient by this symmetry, one requires $\{{\tilde n}_j\}$ to live in the Weyl alcove of $\mathfrak{su}(N-1)$, reducing the cardinality of the set $\{{\tilde n}_j\}$ to that of $\{w_i\}$. Then one can invert the matrix  $\mathfrak{M}_{\{w_i\}, \{{\tilde n}_j\}}$ and obtain the index $C^{T_N}$, which in turn gives the fusion coefficients and the algebra structure of the $SU(N)$ equivariant TQFT.

The metric of the TQFT coming from the cylinder is also straightforward even in the $SU(N)$ case. It is always diagonal and only depends on the symmetry reserved by the holonomy labeled by the highest weight $\lambda$. For instance, if the holonomy is such that $SU(N) \rightarrow U(1)^n \times SU(N_1) \times SU(N_2) \times SU(N_l)$, we have
\beq
\eta^{\lambda {\bar \lambda}} = \frac{1}{(1-\ft)^n} \prod_{j=1}^l \frac{1}{(1-\ft^2)(1-\ft^3)\dots(1-\ft^{N_j})}.
\eeq
This can be generalized to arbitrary group $G$. If the holonomy given by $\lambda$ has stabilizer $G'\subset G$, the norm square of $\lambda$ in the $G_k$ equivariant Verlinde algebra is 
\be
\eta^{\lambda {\bar \lambda}} = P(BG',\ft).
\ee 
Here $P(BG',\ft)$ is the Poincar\'e polynomial\footnote{More precisely, it is the Poincar\'e polynomial in variable $\ft^{1/2}$. But as $H^*(BG,\C)$ is zero in odd degrees, this Poincar\'e polynomial is also a series in $\ft$ with integer powers.} of the infinite-dimensional classifying space of $G'$. In the ``maximal'' case of $G'=U(1)^r$, we indeed get
\be
P\left(BU(1)^r,\ft\right)=P\left(\left(\cp^{\infty}\right)^r,\ft\right)=\frac{1}{(1-t)^r}.
\ee

\appendix
\section{Analytic formula of $\widehat{su}(3)_k$ fusion coefficients}\label{sec: FusionCoeff}

The  notation of this section is from \cite{Begin:1992rt}. Specifically, we define the following quantities:
\beq
k_0^{min}& = \max (\lambda_1 + \lambda_2, \mu_1+\mu_2, \nu_1+\nu_2, a - \min(\lambda_1, \mu_1, \nu_1), b - \min(\lambda_2, \mu_2, \nu_2) ),\\[0.5em]
k_0^{max}& = \min (a,b),
\eeq
where
\beq
a & = \frac{1}{3} \pbra{2(\lambda_1+\mu_1+\nu_1)+\lambda_2+\mu_2+\nu_2},\\[0.5em]
b& = \frac{1}{3} \pbra{\lambda_1+\mu_1+\nu_1+2(\lambda_2+\mu_2+\nu_2)}.\\[0.5em]
\eeq
Moreover we introduce
\be
\delta =  \left \{ \begin{array}{ccc} 1 & \ \ \ \text{if}  \ \ k_0^{max} \geq k_0^{min} & \ \ \text{and} \ \ a, b > 0, a,b \in \mathbb{Z},\\[0.3em]
0 & \text{otherwise}. & \end{array} \right. 
\ee
With these definition we can compactly write our ordinary $su(3)$ representation ring and its fusion coefficient as
\be
N_{\lambda \mu \nu} = (k_0^{max} - k_0^{min} + 1)\delta,
\ee
and we also define a list of $N_{\lambda \mu \nu}$ integers:
\be
{k_0^i} = \{ k_0^{min} , k_0^{min} + 1, \dots, k_0^{max} \}.
\ee
Then the $\widehat{su}(3)_k$ fusion coefficients can be written as
\be
f_{\lambda\mu\nu}(\ft=0) \equiv N_{\lambda\mu\nu}^{(k)} = \left \{ \begin{array}{ccc} \max(i) & \ \ \text{such that}\ \  k > k_0^i \ \ \text{and}\ \ N_{\lambda\mu\nu} \neq 0,  \\[0.5em]
0 & \ \ \ \text{if}\ \  N_{\lambda \mu \nu} = 0 \ \ \text{or} \ \ k < k_0^1. 
\end{array} \right.
\ee

\acknowledgments{We thank J{\o}rgen Ellegaard Andersen, Francesco Benini, Martin Fluder, Abhijit Gadde, Tam\'as Hausel, Murat Kolo\u{g}lu, Pavel Putrov, Richard Wentworth, Ingmar Saberi, Jaewon Song, Andras Szenes and Masahito Yamazaki for helpful discussions related to this work. We would also like to thank the organizers of the Simons Summer Workshop 2015, where a significant portion of this project was completed. This work is funded by the DOE Grant DE-SC0011632, U.S.~National Science Foundation grants DMS 1107452, 1107263, 1107367 (``the GEAR Network''), the Walter Burke Institute for Theoretical Physics, the center of excellence grant ``Center for Quantum Geometry of Moduli Space" from the Danish National Research Foundation (DNRF95), and the Center for Mathematical Sciences and Applications at Harvard University. }

\newpage

\bibliographystyle{JHEP_TD}
\bibliography{draft}

\end{document}